\documentclass[12pt]{article}
\pdfoutput=1
\usepackage{epsf,amsmath,amssymb,graphicx,dcolumn,multirow}
\usepackage{scalefnt,cite,hyperref}
\usepackage{cleveref,etoolbox,color,soul}
\usepackage{listings}
\usepackage{xspace}
\usepackage[utf8]{inputenc}
\usepackage{booktabs, colortbl}
\usepackage{cancel}

\Crefname{figure}{Fig.}{Figs.}
\usepackage{placeins}
\usepackage{soul}

\newcommand{\citere}[1]{Ref.\,\cite{#1}}
\newcommand{\citeres}[1]{Refs.\,\cite{#1}}
\newcommand{\code}{\tt}

\newcommand{\sushi}{{\code SusHi}\xspace}
\newcommand{\FH}{{\code FeynHiggs}\xspace}
\newcommand{\HiggsB}{{\code HiggsBounds}\xspace}
\newcommand{\HiggsS}{{\code HiggsSignals}\xspace}

\newcommand{\abbrev}{\scalefont{1}}
\newcommand{\eqn}[1]{Eq.\,(\ref{#1})}
\newcommand{\eqns}[2]{Eqs.\,(\ref{#1}) and (\ref{#2})}
\newcommand{\fig}[1]{Fig.\,\ref{#1}}
\newcommand{\figs}[1]{Figs.\,\ref{#1}}
\newcommand{\tab}[1]{Tab.\,\ref{#1}}
\newcommand{\sct}[1]{Section~\ref{#1}}

\newcommand{\lhc}{{\abbrev LHC}\xspace}

\newcommand{\eft}{{\abbrev EFT}\xspace}
\newcommand{\sm}{{\abbrev SM}\xspace}
\newcommand{\thdm}{{\abbrev 2HDM}\xspace}
\newcommand{\thdmew}{{\abbrev 2HDM+EWinos}\xspace}
\newcommand{\mssm}{{\abbrev MSSM}\xspace}

\newcommand{\lsp}{{\abbrev LSP}}
\newcommand{\susy}{{\abbrev SUSY}}
\newcommand{\bsm}{{\abbrev BSM}}
\newcommand{\pdf}{{\abbrev PDF}}

\newcommand{\lhchxswg}{{\abbrev LHC-HXSWG}\xspace}

\newcommand{\cp}{{\abbrev $\mathcal{CP}$}}
\newcommand{\nlo}{{\abbrev NLO}}
\newcommand{\nnlo}{{\abbrev NNLO}}
\newcommand{\nklo}[1]{{\abbrev N$^{#1}$LO}}
\newcommand{\nll}{{\abbrev NLL}}
\newcommand{\drbar}{{\abbrev $\overline{\text{DR}}$}}

\newcommand{\GeV}{{\rm GeV}}
\newcommand{\TeV}{{\rm TeV}}

\newcommand{\mW}{\ensuremath{M_W}}
\newcommand{\mZ}{\ensuremath{M_Z}}
\newcommand{\mh}{\ensuremath{M_h}}

\newcommand{\mA}{\ensuremath{M_A}}

\newcommand{\msusy}{\ensuremath{M_{\text{\susy}}}\xspace}
\newcommand{\tb}{\ensuremath{\tan\beta}}

\definecolor{darkgreen}{rgb}{0,0.5,0.1}

\newcommand{\etmiss}{\cancel{E}_T}

\newcommand{\mhsc}{$\mh^\text{125}$\xspace}

\newcommand{\lchi}{$\mh^{125}(\tilde{\chi})$\xspace}

\newcommand{\ltbmhsc}{$M_{h,\text{\eft{}}}^\text{125}$\xspace}
\newcommand{\ltblchi}{$M_{h,\text{\eft{}}}^{125}(\tilde{\chi})$\xspace}
\newcommand{\HiBo}{\texttt{HiggsBounds}}
\newcommand{\HiSi}{\texttt{HiggsSignals}}

\definecolor{Gray}{gray}{0.9}
\definecolor{DGray}{gray}{0.7}

\begin{document}
\thispagestyle{empty}
\begin{flushleft}
\today
\end{flushleft}
\begin{flushright}
{\tt DESY 18-219},
{\tt KA-TP-01-2019}
\end{flushright}

\long\def\symbolfootnote[#1]#2{\begingroup%
\def\thefootnote{\fnsymbol{footnote}}\footnote[#1]{#2}\endgroup}

\vspace{0.1cm}

\begin{center}
\Large\bf\boldmath
MSSM Higgs Benchmark Scenarios for\\ Run 2 and Beyond: the low $\tb$ region
\unboldmath
\end{center}
\vspace{0.05cm}
\begin{center}
Henning Bahl$^a$,
Stefan Liebler$^b$,
Tim Stefaniak$^{a}$\\[0.4cm]
{\small
{\sl${}^a$DESY, Notkestra{\ss}e 85, D-22607 Hamburg, Germany}\\[0.2em]
{\sl${}^b$Institute for Theoretical Physics (ITP), Karlsruhe Institute of Technology,\\ D-76131 Karlsruhe, Germany}
}
\end{center}
\vspace*{1mm}
\begin{abstract}
\noindent

We propose two new benchmark scenarios for Higgs-boson searches in the Minimal Supersymmetric Standard Model (\mssm{}). These scenarios are specifically designed for the low $\tb$ region. A light Higgs-boson mass prediction compatible with the observed value of $125$\,GeV is ensured in almost the entire parameter space by employing a flexible supersymmetric (\susy{}) mass scale, reaching values of up to $10^{16}$\,GeV. The \mssm{} Higgs-sector predictions are evaluated in an effective field theory (\eft{}) framework that exhibits a Two-Higgs-Doublet-Model at the low scale. In the first scenario all \susy{} particles are relatively heavy, whereas the second scenario features light neutralinos and charginos. Both scenarios are largely compatible with the most recent results from Run 2 of the \lhc{}, and we highlight the main phenomenological features relevant for future \lhc{} searches. In particular, we provide a detailed discussion of heavy Higgs-boson decays to neutralinos and charginos in the second scenario, and the arising collider signatures, in order to facilitate the design of dedicated \lhc{} searches in the near future.

\end{abstract}

\setcounter{footnote}{0}

\newpage

\pagenumbering{arabic}


\section{Introduction}
\label{sec:introduction}
The last free parameter in the Standard Model (SM) of particle physics, namely the mass of the Higgs boson that was discovered at the Large Hadron Collider (\lhc)~\cite{Aad:2012tfa,Chatrchyan:2012xdj}, was determined during Run~1 of the \lhc{} to $M_{H^{\text{\sm{}}}}^{\rm obs} = 125.09 \pm 0.24$\,GeV~\cite{Aad:2015zhl}. In addition, the Higgs-boson properties, among them the Higgs-boson couplings to the heavier \sm{} particles and the Higgs-boson width, were found to be compatible with the \sm{} predictions within the experimental and theoretical uncertainties~\cite{Khachatryan:2016vau}. This experimental information puts strong indirect constraints on beyond-the-\sm{} (\bsm{}) physics, which become increasingly important since no direct evidence for \bsm{} physics has yet been found at the \lhc{}.

One of the most compelling models of \bsm{} physics is the Minimal Supersymmetric Standard Model (\mssm{})~\cite{Nilles:1983ge,Haber:1984rc,Gunion:1984yn}. Apart from associating a superpartner with each \sm{} degree of freedom, it also extends the \sm{} Higgs sector by a second doublet resulting in five physical Higgs states. Assuming \cp{} conservation in the Higgs sector these are the light and heavy \cp-even Higgs bosons, $h$ and $H$, respectively, the \cp-odd Higgs boson, $A$, and the charged Higgs bosons, $H^\pm$.
Due to the underlying supersymmetry (\susy{}), the \mssm{} Higgs sector is highly predictive. At the tree-level, all masses and couplings are determined by only two non-\sm{} parameters. Typically, the ratio of the vacuum expectation values (vevs) of the two doublets, $\tb$, and the mass of the \cp-odd Higgs boson, $M_A$, are chosen in the \cp-conserving \mssm.

However, the tree-level predictions receive large quantum corrections. Their inclusion is especially important in the case of the \sm{}-like Higgs-boson mass, which we assume to be the $h$~boson in this work (see \citeres{Heinemeyer:2011aa,Drees:2012fb,Bechtle:2012jw, Bechtle:2016kui,Haber:2017erd} for scenarios in which the $H$~boson plays the role of the Higgs boson discovered at the \lhc{}). Consequently, much work has been dedicated to the calculation of these corrections. We refer to \citere{Draper:2016pys} for a recent review.
Through the radiative corrections a large set of parameters enter the calculation. This makes an interpretation of corresponding experimental results challenging. Consequently, benchmark scenarios have been proposed to alleviate experimental analyses and their theoretical interpretation~\cite{Carena:1999xa, Carena:2002qg,Carena:2013ytb,Carena:2000ks,Bagnaschi:2039911,Carena:2015uoe}. Due to experimental and theoretical progress much of the parameter space of these original benchmark scenarios has been ruled out. Therefore, new benchmark scenarios, taking into account the most recent experimental limits as well as state-of-art theory predictions, have been proposed in~\citere{Bahl:2018zmf}.

In all the original benchmark scenarios as well as in the new scenarios presented in \citere{Bahl:2018zmf} the supersymmetric partners of the \sm{} fermions (sfermions) are tied to the TeV scale. In this case, the parameter region $\tb\lesssim 5$ is ruled out because the mass of the \sm{}-like Higgs boson, $M_h$, is predicted to be lower than the measured value. On the other hand, experimental searches for heavy Higgs bosons decaying into bottom quarks and $\tau$ leptons~\cite{Aaboud:2017sjh,Sirunyan:2018zut} rule out the parameter space at large $\tb$ up to high values of $M_A$ beyond the TeV scale. At low $\tb$ the searches for heavy Higgs bosons are much less sensitive, i.e. still allow for lower values of $M_A$. In this region, the Higgs bosons decay into a variety of final states, which are more difficult to handle experimentally. To re-open the parameter region of low $\tb$ values, the ``low-$\tb$-high'' scenario was proposed in \citere{Bagnaschi:2039911} raising the sfermion mass scale, $\msusy$, up to $100$~TeV in order to reach $M_h\sim 125$\,GeV also for low $\tb$. In case of such a large hierarchy between the electroweak scale and the sfermion scale, large logarithms, involving $\msusy$ and e.g. the top-quark mass, appear in the calculation of $M_h$. These can be resummed by integrating out the sfermions at $\msusy$ and then evolving the couplings in the effective field theory (\eft{}) below \msusy using renormalization group equations (RGEs) down to the electroweak scale at which the \sm{}-like Higgs-boson mass is calculated. In the simplest approach the \sm{} is used as \eft{} \cite{Giudice:2011cg,Hahn:2013ria,Draper:2013oza,Bagnaschi:2014rsa,Vega:2015fna,Bahl:2016brp,Bagnaschi:2017xid,Bahl:2017aev,Harlander:2018yhj}. Also the ``low-$\tb$-high'' scenario was based on this setup. However, if we assume all Higgs-boson masses to be close to the electroweak scale, i.e. below a few TeV, the Two-Higgs-Doublet Model (\thdm{}) has to be used as low-energy \eft{} to ensure a correct resummation. Such setups were published in  \citeres{Haber:1993an,Lee:2015uza,Athron:2017fvs,Bahl:2018zmf,Gabelmann:2018axh}. Those works show that a substantial part of the parameter space in the ``low-$\tb$-high'' scenario yields a prediction for the light Higgs-bosons mass, $M_h$, which is much lower than the measured value and hence the ``low-$\tb$-high'' scenario is meanwhile ruled out. It is the goal of this paper to define new scenarios valid at low values of $\tb$ in the framework of a low-energy \thdm{}.

Moreover, this region of low $\tb$ is the region of validity of the hMSSM~\cite{Djouadi:2013vqa,Maiani:2013hud,Djouadi:2013uqa}, an approximation of the \mssm{} Higgs sector, which assumes that the dominant correction to Higgs boson masses and mixing have a common origin: they stem from the top-quark and its \susy{} partners, the stops, entering a single element of the neutral \cp{}-even Higgs-boson mass matrix only. As indicated this assumption is only valid at low values of $\tb$ and in addition for low values of $\mu/\msusy$, and then allows to trade the loop corrections as a function of the light Higgs boson mass $M_h$ (for specific values of $M_A$ and $\tb$) such that the explicit dependence of the Higgs masses and mixing on the \susy{} parameters can be ignored. For low $\tb$ the original hMSSM approach is known to approximate Higgs masses and mixing quite well, see e.g. \citeres{Djouadi:2015jea,Bagnaschi:2039911,Lee:2015uza}, though for the Higgs-boson couplings, entering e.g. the important decay $H\to hh$, further refinements~\cite{Liebler:2018zul} are needed. In comparison with our suggested scenarios the quality of the hMSSM approximation and its region of validity can be tested. As a first step, we compare the \cp{}-even Higgs-boson mixing angle, $\alpha$, and the heavy \cp{}-even Higgs-boson mass, $M_H$, between the two approaches, but leave a discussion of more elaborate quantities, including the prediction of the $H\to hh$ partial width, to future work.

We shall define two scenarios: One scenario in which all non-\thdm states are decoupled and one featuring light neutralinos and charginos, to which we refer as electroweakinos. The first scenario resembles the \mhsc scenario, the other the \lchi scenario, which were defined in \citere{Bahl:2018zmf}. In order to allow for $M_h\sim 125$\,GeV also at low $\tb$, a very high sfermion mass scale of up to $10^{16}$\,GeV is needed.
We employ a state-of-the-art calculation of the Higgs-boson masses and branching ratios using a yet unpublished version of \FH~\cite{Heinemeyer:1998yj, Heinemeyer:1998np,Degrassi:2002fi,Frank:2006yh, Hahn:2013ria, Bahl:2016brp,Bahl:2017aev,Bahl:2018qog}. This version implements the results of \citere{Bahl:2018jom}: The effective \thdm (and a \thdmew) as \eft{} below the sfermion scale includes full two-loop renormalization group equations as well as full one-loop and partial two-loop threshold corrections. This \eft{} calculation is combined with a state-of-the-art fixed-order calculation. For more details we refer to \citere{Bahl:2018jom}. We obtain the production cross sections of the neutral Higgs bosons from \sushi~\cite{Harlander:2012pb,Harlander:2016hcx} for gluon fusion and re-weight matched predictions
published in \citeres{Forte:2016sja,Forte:2015hba,Bonvini:2016fgf,Bonvini:2015pxa} for bottom-quark associated production.
We explore the known experimental constraints on the \mssm{} Higgs sector using \HiggsB~\cite{Bechtle:2008jh,Bechtle:2011sb,Bechtle:2013wla,Bechtle:2015pma} and \HiggsS~\cite{Bechtle:2013xfa}.

The paper is set up as follows: We explain our theoretical setup in detail in \sct{sec:theorysetup}. We proceed with an explanation of the experimental constraints in \sct{sec:exp}.
Finally we define and discuss our two benchmark scenarios in \sct{sec:benchmarks} including a first comparison with the hMSSM approach and conclude thereafter.


\section{Theory setup}
\label{sec:theorysetup}
In this Section, we provide details about the calculation of the
Higgs-boson masses and branching ratios as well as of the Higgs-boson
production cross sections.


\subsection{Higgs-boson masses and branching ratios}
For the calculation of the Higgs-boson properties (masses and branching ratios), we rely on the code \FH. Since the scenarios presented in this work include a large hierarchy between the scale of non-\sm{}-like Higgs bosons and scalar fermions, we use a yet not public \FH version\footnote{Until its official release the employed version of FeynHiggs can be obtained from the authors upon request.} which implements a low-energy \thdm as well as a low-energy \thdmew as effective field theories \cite{Bahl:2018jom},\footnote{We corrected the renormalization group equations used in \citere{Bahl:2018jom} according to the findings of \citeres{Bednyakov:2018cmx,Schienbein:2018fsw}. The numerical effect is negligibly small.} which is merged with a one-loop diagrammatic fixed-order calculation \cite{Chankowski:1992er,Dabelstein:1995js,Pierce:1996zz,Frank:2006yh}. The two-loop diagrammatic fixed-order corrections implemented in \FH are switched off, see below.

For future reference, we list the \FH flags set to obtain the result presented in this work,
\vspace*{-6mm}
\begin{center}
\text{mssmpart} = 4,
\quad \text{higgsmix} = 2,\quad \text{p2approx} = 4,\quad
 \text{looplevel} = 1,\\[2mm]
\text{loglevel} = 4,\quad \text{runningMT} = 1,
\quad \text{botResum} = 1,\quad \text{tlCplxApprox} = 0.
\end{center}
Setting loglevel~=~4 enables the \thdm as low-energy theory. This flag choice is not yet available in the most recent public \FH version {\code 2.14.3}, but will become available in an upcoming version.

In order to provide a numerical stable prediction for very high \susy{} scales of up to $10^{16}$\,GeV, the code has to be compiled using quadruple precision. To improve numerical stability, also an improved solving algorithm for the renormalization group equation was implemented. In addition, we have to deactivate the diagrammatic two-loop corrections to the Higgs-boson self-energies, which are numerically unstable in this parameter region even in case of quadruple precision. Note that two-loop terms which are not suppressed by the \susy{} scale, re-enter via the \eft{} calculation.\footnote{This includes corrections involving only \thdm particles.}  The missing suppressed terms are completely negligible. For the same reason, the used one-loop Higgs-boson self-energies are expanded in the limit of large sfermion masses. We want to point out that with this \FH configuration $\tb$ is defined in the \thdm at the scale $M_A$. In contrast, in the benchmark scenarios defined in \citere{Bahl:2018zmf}, $\tb$ is evaluated at the scale of the top-quark mass, $M_t$.

We outlined the history of \eft{} calculations of the Higgs-boson mass already in the introduction. Apart from \FH there are two other codes implementing the \thdm{} as low-energy \eft{}. These are {\code MhEFT}~\cite{Lee:2015uza} and~{\code FlexibleSUSY} \cite{Bagnaschi:2015pwa,Athron:2017fvs}.\footnote{They do provide predictions for the Higgs-boson masses but not (yet) for their branching ratios.} Among these codes, the prediction of the light Higgs-boson mass, $M_h$, is in good agreement. A comparison of these codes with \FH, however, yields discrepancies of a few GeV in the prediction of $M_h$ in the region of $M_A \sim 200$\,GeV, $\tb\sim 1$, $\mu \sim 200$\,GeV and \msusy above $10^{10}\;\GeV$, where \FH predicts higher $M_h$ values.\footnote{In contrast to the other codes, the EFT calculation implemented in \FH\ takes into account all effective couplings as well as full one-loop threshold corrections, see \citere{Bahl:2018jom} for more details. This may be the origin of the observed numerical discrepancies.} Thus using the other codes instead could exclude parts of this region of the parameter space by a too low prediction of $M_h$. As we will show in \sct{sec:benchmarks}, this region is already excluded by the Higgs-boson signal-strength measurements.
For higher $\tb$ or $M_A$ values $\msusy$ can be adjusted to get $M_h\sim 125$\,GeV without changing the low-energy phenomenology. Therefore, while it will be still important to understand the observed discrepancy between the \FH\ and {\code MhEFT}/{\code FlexibleSUSY} predictions, the outcome of this discussion will hardly affect the phenomenology of the scenarios presented in this work.


\subsection{Production cross sections}
\label{sec:prod}
For the calculation of the gluon-fusion cross sections of the neutral Higgs bosons we employ {\code SusHi 1.7.0} \cite{Harlander:2012pb,Harlander:2016hcx}, which we directly link to \FH at quadruple precision. We take into account the Higgs-boson mixing by implementing the full $\mathbf{Z}$ matrix as explained in \citere{Liebler:2016ceh}. The $\mathbf{Z}$ matrix relates the tree-level mass eigenstates to the external physical states and is calculated by \FH including a resummation of large logarithms by the means of the EFT calculation \cite{Frank:2006yh,Fuchs:2016swt,Bahl:2018ykj}. \sushi{} includes top- and bottom-quark contributions at next-to-leading order (\nlo{})~\cite{Spira:1995rr,Harlander:2005rq} and for the top-quark contribution adds next-to-\nlo{} (\nnlo) effects in the heavy-quark effective theory~\cite{Harlander:2002wh,Anastasiou:2002yz,Ravindran:2003um,Harlander:2002vv,Anastasiou:2002wq} as well as next-to-NNLO (\nklo{3}) contributions, which additionally exploit a threshold expansion~\cite{Anastasiou:2014lda, Anastasiou:2015yha,Anastasiou:2016cez}. The latter \nklo{3} contributions are only taken into account for the light \sm{}-like Higgs boson to match the precision of the \lhchxswg\ gluon-fusion cross section in the \sm{}, see \citere{deFlorian:2016spz}, except from \nnlo{} top-quark mass effects. The \nklo{3} contribution in the threshold expansion closely matches the exact result published in \citere{Mistlberger:2018etf}. We also include two-loop electroweak corrections from light quarks as discussed in \citeres{Aglietti:2004nj,Bonciani:2010ms}. On the other hand, we omit corrections from \susy{} particles to gluon fusion, both explicit in the amplitude and through $\Delta_b$ corrections. This is justified from the fact that those contributions quickly decouple for high \susy{} masses as employed in this work. Their inclusion would induce numerical instabilities due to the high \susy{} masses. Keeping $\mu$ at the electroweak scale and fixing the gluino mass, $M_3$, to $2.5$\,TeV, this statement also applies to the dominant $\Delta_b$ corrections. We make use of the parton distribution functions (\pdf{}) named {\verb=PDF4LHC15_nlo_mc=} and {\verb=PDF4LHC15_nnlo_mc=}~\cite{Butterworth:2015oua} and choose half of the Higgs-boson mass as central renormalization and factorization scale.

Instead of employing ``Santander-matched'' cross sections~\cite{Harlander:2011aa} for the production of neutral Higgs bosons
through bottom-quark annihilation, we follow the recommendation of the \lhchxswg{}~\cite{deFlorian:2016spz}
and use the results based on
soft-collinear effective theory~\cite{Bonvini:2015pxa, Bonvini:2016fgf} and the
``fixed order plus next-to-leading log'' (FONLL) approach~\cite{Forte:2015hba, Forte:2016sja},
which both yield identical cross sections. They are based on the cross sections
obtained in the five-flavor scheme~\cite{Harlander:2003ai} and in the four-flavor scheme~\cite{Dittmaier:2003ej,Dawson:2003kb}.
We re-weight the cross section of the \sm{} Higgs boson proportional to the squared bottom-quark Yukawa coupling both for the \cp{}-even and \cp{}-odd Higgs bosons
and omit corrections proportional to the top-quark Yukawa coupling.

For a detailed discussion of the theoretical uncertainties of neutral Higgs-boson production
in the \mssm{} we refer to \citere{Bagnaschi:2014zla}.
We obtain theoretical uncertainties for gluon fusion
and bottom-quark annihilation in the same way as discussed in \citere{Bahl:2018jom}, in summary:
For gluon fusion we include the renormalization-scale uncertainty, which is calculated analytically
from $100$ scale choices between half and twice the central scale choice following the approach discussed in \citere{Harlander:2016hcx}.
The difference between the maximal and minimal cross section is used as symmetric uncertainty. The factorization-scale uncertainty is subdominant and not further considered.
We take into account relative \pdf{} and $\alpha_s$ uncertainties obtained as a function of the
Higgs-boson mass for a \sm{} Higgs boson and for a \thdm{} \cp{}-odd Higgs boson at $\tb=1$, which we employ for \cp{}-even and the \cp{}-odd Higgs bosons, respectively.
The absolute renormalization-scale and \pdf{}$+\alpha_s$ uncertainties are added in quadrature.
For bottom-quark associated production we use the absolute uncertainties provided by the \lhchxswg{} both for \cp{}-even and \cp{}-odd Higgs bosons,
which include renormalization- and factorization-scale uncertainties, uncertainties related to the value of the bottom-quark mass and the bottom-quark matching scale
and \pdf{}$+\alpha_s$ uncertainties.

Lastly we employ cross sections for charged Higgs-boson production according to the recommendation of the
\lhchxswg{}, which is based upon \citeres{Berger:2003sm, Dittmaier:2009np,Flechl:2014wfa,Degrande:2015vpa, Degrande:2016hyf}.
This in particular includes cross sections for charged Higgs bosons in the mass window of $145$--$200$\,GeV
at a center-of-mass energy of $13$\,TeV.

\section{Experimental constraints}
\label{sec:exp}
Searches for additional Higgs bosons at the \lhc{}, as well as the signal-strength measurements of the observed Higgs boson at the \lhc{}, already constrain part of the parameter space of our benchmark scenarios. These constraints are tested in our analysis with the codes \HiBo~\cite{Bechtle:2008jh,Bechtle:2011sb,Bechtle:2013wla,Bechtle:2015pma}  and \HiSi~\cite{Bechtle:2013xfa}, respectively. We follow the same procedure as in Ref.~\cite{Bahl:2018zmf}, which we summarize briefly in the following.

\subsection{Constraints from LHC searches for additional Higgs bosons}
\label{sec:HB}
The program \HiBo\ tests each parameter point against the $95\%~\mathrm{C.L.}$ cross-section limits from neutral and charged Higgs boson searches at the LEP, Tevatron and \lhc{} experiments. It turns out, however, that only the \lhc{} Higgs searches are important in our benchmark scenarios. The code follows a well-defined statistical procedure when applying these constraints: in a first step, it determines the most sensitive experimental search for each Higgs boson in the model (as judged by the expected limit); in the second step only the observed upper limit of this most sensitive search is compared to the model-predicted signal rate, and the model is regarded as excluded if the predicted rate exceeds the upper limit. We refer to \citeres{Bechtle:2008jh,Bechtle:2011sb,Bechtle:2013wla,Bechtle:2015pma} for more~details.

The latest version of \HiBo, {\tt 5.3.0beta}, includes results from the following \lhc{} searches relevant to our scenarios: searches for heavy Higgs bosons decaying to $\tau^+\tau^-$ pairs by ATLAS~\cite{Aaboud:2017sjh} and CMS~\cite{Sirunyan:2018zut} using about $36\,\mathrm{fb}^{-1}$ of Run-2 data, as well as the CMS results from Run~1~\cite{CMS:2015mca}; ATLAS~\cite{Aad:2015kna, Aaboud:2017rel} and CMS~\cite{Khachatryan:2015cwa, Sirunyan:2018qlb} searches during Run-1 and Run-2 for a heavy scalar decaying to a $Z$-boson pair; Run-2 searches by ATLAS~\cite{Aaboud:2018sfw} and CMS~\cite{Sirunyan:2017djm, Sirunyan:2017guj,CMS:2017orf,Sirunyan:2018ayu} for a heavy scalar decaying to a pair of \sm{}-like Higgs bosons. The most relevant charged Higgs-boson searches included in \HiBo\ are searches for top-quark associated $H^\pm$ production, with subsequent decays to $\tau\nu$~\cite{Khachatryan:2015qxa, CMS:2016szv, Aad:2014kga, Aaboud:2018gjj} or $tb$~\cite{Khachatryan:2015qxa, Aad:2015typ, Aaboud:2018cwk} pairs.

We estimate the theoretical uncertainty in our determination of the excluded regions by re-evaluating the excluded region with \HiBo\ for (\emph{i}) a most conservative and (\emph{ii}) a least conservative variation of the gluon-fusion and bottom-quark annihilation cross sections by their estimated uncertainties, see \sct{sec:prod} and \citere{Bahl:2018jom}.

\subsection{Constraints from the Higgs boson observed at the LHC}
We test the compatibility with the observed Higgs-boson signal rates with the program \texttt{Higgs\-Signals}~(version \texttt{2.2.1beta}). The program includes the combined \lhc{} Run-1 ATLAS and CMS measurements of the Higgs-boson signal strengths~\cite{Khachatryan:2016vau} as well as recent measurements during \lhc{} Run-2 that became available in mid-2018, with around $36~\mathrm{fb}^{-1}$ of integrated luminosity per experiment~\cite{ATLAS:2016gld, ATLAS:2018gcr, Aaboud:2017xsd, Aaboud:2017vzb, Aaboud:2017jvq, Aaboud:2017rss, Aaboud:2018xdt,CMS:2017rli, Sirunyan:2017exp, Sirunyan:2017khh, Sirunyan:2017dgc, Sirunyan:2017elk, Sirunyan:2018shy, Sirunyan:2018ygk, Sirunyan:2018mvw, Sirunyan:2018egh,  Sirunyan:2018hbu}. The program evaluates a $\chi^2$ value for each parameter point using in total $100$ individual signal-rate measurements. Using this $\chi^2$ value, we perform an (approximate) log-likelihood ratio test for model discrimination as follows: within the considered two-dimensional benchmark plane, we determine the parameter point with minimal $\chi^2$ value, $\chi^2_\text{min}$, and regard all parameter points with $\Delta\chi^2 \equiv \chi^2 - \chi^2_\text{min} \le 6.18$ to be consistent at the $2\sigma$ level with the best-fit hypothesis (and, hence, with the observed Higgs rates). We remark that, in both of our proposed benchmark scenarios, the best fit point is found far in the decoupling limit ($M_A \gg M_Z$), where the model provides essentially as good a fit to the observed Higgs rates as the \sm{}.


\section{Benchmark scenarios}
\label{sec:benchmarks}
We subsequently explain our parameter choices and present two benchmark scenarios for low values of $\tb$ between $1$ and $10$, based on the previously discussed \eft{} setup employing a very heavy colored \susy{} spectrum.
Both are inspired by two scenarios presented in \citere{Bahl:2018jom}, which work with a TeV-scale colored \susy{} spectrum
and are thus only valid at higher values of $\tb$.


\subsection{Input parameters}
Following the recommendation of the \lhchxswg in \citere{deFlorian:2016spz}, we make use of the following \sm input parameters:
\begin{eqnarray}
&m_t^{\text{pole}}=172.5~\text{GeV},\quad
\alpha_s(\mZ)=0.118,\quad
G_F=1.16637\cdot 10^{-5}~\text{GeV}^{-2},\nonumber\\
&m_b(m_b)=4.18~\text{GeV},\quad
\mZ=91.1876~\text{GeV},\quad
\mW=80.385~\text{GeV}\,.
\end{eqnarray}
The other lepton and quarks masses only have a minor influence on the Higgs-sector observables. Therefore, we stick to the default values of \FH. There, however, is a strong dependence of the light Higgs-boson mass, $M_h$, on the value of the employed top-quark pole mass. The value recommended by the \lhchxswg is below the current world average of $173.21\pm 0.51\pm 0.71$\,GeV~\cite{Olive:2016xmw}. In the scenarios considered here, a change of the top-quark pole mass of $0.7\;\GeV$ would imply a change of $\sim 0.8\;\GeV$ in the prediction for $M_h$.

To fix the \susy{} parameters, we choose to set all scalar fermion soft-\susy{} breaking masses equal to a common scale $\msusy$. In both scenarios, presented here, $\msusy$ will be adjusted at each point in the ($M_A$, $\tb$) plane such that $M_h\sim 125$\,GeV is reached.\footnote{Tables listing the $\msusy{}$ values are available as auxiliary material to this manuscript.} We, however, do not allow $\msusy$ to be larger than $10^{16}$\,GeV. The minimal value of $\msusy$ in both benchmark scenarios is $\sim 6\;\TeV$, when restricting to values of $\tb$ between $1$ and $10$ and $M_A$ to be less than $2$\,TeV.

In addition, in both scenarios, we choose for the gluino mass and the third-generation soft-\susy{} breaking trilinear couplings,
\begin{eqnarray}
 M_3=2.5~\text{TeV} \,,\quad
A_t=A_b=A_\tau=0\,,
\label{eq:commonpara}
\end{eqnarray}
respectively.
Typically, large $A_t$ values are chosen to reach $M_h\sim 125$\,GeV for low $\msusy$ values. In the scenarios considered here, we are interested most in the region of low $\tb$ and low $M_A$. In this region, very high $\msusy$ values of up to $10^{16}$\,GeV are needed to reach $M_h\sim 125$\,GeV. For such high values of $\msusy$, there is only a mild dependence of the prediction for $M_h$ on the size of the stop mixing. And due to theoretical fine-tuning arguments, low $A_t$ values are preferred in case of a TeV-scale $M_3$ and TeV-scale electroweakinos.
Fixing $M_3$ to $2.5$\,TeV, the gluino mass is safely above current bounds from direct searches~\cite{Aaboud:2017bac,Aaboud:2017vwy,Sirunyan:2017cwe,Sirunyan:2017kqq,Sirunyan:2018vjp,Aaboud:2017dmy}.


\subsection{\texorpdfstring{$\boldsymbol{M}_{\boldsymbol{h},\text{\eft{}}}^{\text{125}}$}{MhEFT125} scenario}
The first benchmark scenario we propose is the \ltbmhsc scenario. All \susy{} particles are chosen to be heavy. Consequently, all \mssm{} Higgs boson collider observables are only mildly affected by \susy{} particles and the phenomenology is very similar to that of a type-II~\thdm. This scenario serves as a phenomenologically viable extension of the \mhsc scenario presented in \citere{Bahl:2018zmf} to low $\tb$ values\footnote{Low values of $\tan\beta$ in an effective \thdm{} are also well motivated from flavor and stability constraints~\cite{Bhattacharyya:2017ksj}.}. Hence, we choose the same Higgsino, bino and wino mass parameters,
\vspace*{-4mm}
\begin{eqnarray}
&  \mu=1~\text{TeV},\,\quad
M_1=1~\text{TeV},\quad M_2=1~\text{TeV},
\label{eq:ltbmhscscenario}
\end{eqnarray}
respectively. The other input parameters are fixed according to \eqn{eq:commonpara}. With this parameter choice, the scenario is similar to the old ``low-$\tb$-high'' scenario~\cite{Bagnaschi:2039911}, where the electroweakinos were also chosen to have masses around the TeV scale. The \ltbmhsc scenario is a concrete realization of an \mssm{} scenario, which fulfills the assumptions used in the hMSSM approach~\cite{Djouadi:2013vqa,Djouadi:2013uqa,Djouadi:2015jea}: It is defined in the region of interest, i.e. low $\tb$ and low $M_A$, where for low $\mu/\msusy{}$ the dominant corrections to the Higgs-boson mass matrix stem from a single element in the $(2\times 2)$ \cp{}-even Higgs-boson mass matrix. Therefore it is a perfect candidate for a more detailed comparison of remaining discrepancies in Higgs-boson mass and mixing predictions as well as Higgs-boson self-couplings and Higgs-to-Higgs decays, see e.g. \citere{Bagnaschi:2039911}. Those discrepancies could reveal potential limitations of the hMSSM approach.
We present a comparison of the heavy \cp{}-even Higgs-boson mass, $M_H$, and the Higgs-boson mixing angle, $\alpha$, in the following subsection.

\begin{figure}[t]
\begin{center}
\vspace*{-4mm}
\includegraphics[width=0.75\textwidth]{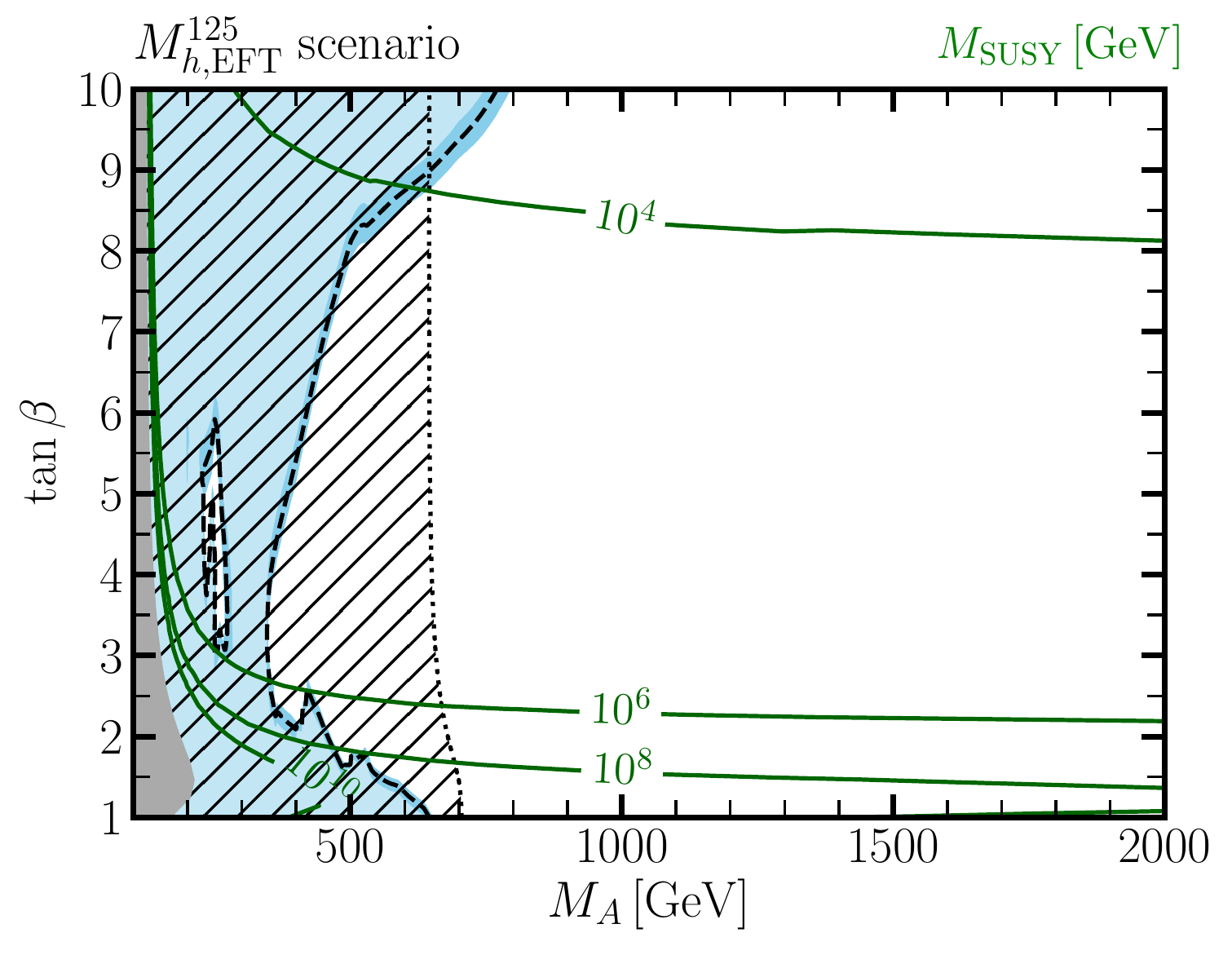}
\end{center}
\vspace*{-10mm}
\caption{The \ltbmhsc benchmark scenario shown in the $(\mA\,,\tb)$ plane. The \emph{green solid contour lines} indicate the required \susy{} scale. The \emph{black dotted line} and \emph{hatched area} marks the parameter space which is disfavored at the $2\sigma$ level by the measured Higgs-boson signal rates. The \emph{blue region} with the \emph{black dashed edge} is excluded at the $95\%~\mathrm{C.L.}$ by \lhc{} searches for additional Higgs bosons (the \emph{dark blue band} shows how the theoretical rate uncertainty affects the exclusion). The \emph{gray area} is excluded because the mass of the \sm{}-like Higgs boson $h$ is below $122$\,GeV.}
\label{fig:mh125-ltb_msusy}
\end{figure}

In \fig{fig:mh125-ltb_msusy} we present the current constraints on the \ltbmhsc scenario in the ($M_A$, $\tb$) parameter plane. As described above, the \susy{} mass scale, \msusy, is adjusted at every point in order to obtain $M_h\simeq 125$\,GeV throughout the parameter plane. In the \emph{gray area}, however, $M_h < 122$\,GeV, since \msusy would have to be raised above our imposed upper limit of $10^{16}$\,GeV in order to obtain large $M_h$ values.\footnote{Outside of the \emph{gray region}, the prediction for $M_h$ quickly increases from $M_h = 122$\,GeV to $M_h = 125$\,GeV.} This part of the parameter space is therefore excluded.

Direct searches at the \lhc{} for non-\sm{}-like Higgs bosons exclude the \emph{blue region} in \fig{fig:mh125-ltb_msusy}, as determined with \HiBo. The \emph{dark blue band} indicates the theoretical uncertainty of the exclusion, obtained by varying the cross section of $H$ and $A$ production in gluon fusion and in association with a bottom-quark pair, as described in \sct{sec:prod} and \sct{sec:HB}. For $\tb$ values larger than around $3$--$4$, the most important search channel is $gg/b\bar{b}\to H/A \to \tau^+\tau^-$. The CMS analysis~\cite{Sirunyan:2018zut} is slightly more sensitive than the ATLAS analysis~\cite{Aaboud:2017sjh} and excludes $M_A$ values of $\lesssim 400~(750)$\,GeV for $\tb \sim 5~(10)$, except for a small parameter region at $M_A \sim 250$\,GeV and $\tb \sim 3 - 6$. In this region, \lhc{} searches for $pp\to H \to ZZ$ show similar or even higher sensitivity. The applied ATLAS search limit~\cite{Aaboud:2017rel}, however, exhibits some statistical fluctuations in this region, leading to the observed hole (and tiny spike) in the exclusion.\footnote{Note that a proper combination of experimental results by the \lhc{} collaborations would presumably close the gap in the exclusion.} At lower $\tb$ values direct searches for $pp\to H\to hh$ become important. Currently, the most sensitive final state of this process is $\tau^+\tau^- b\bar{b}$, with the CMS search~\cite{CMS:2017orf} being more sensitive for lower masses, $M_A < 420$\,GeV, and the ATLAS search~\cite{Aaboud:2018sfw} for higher masses up to $M_A \sim 500$\,GeV.\footnote{The transition between the two analyses results in the spike observed in the exclusion region at ${M_A \sim 420}$\,GeV and ${\tb\sim 2.5}$.} Beyond $M_A \sim 500$\,GeV, at very low $\tb$ values, the exclusion arises from the CMS combination of $H\to hh \to b\bar{b}\gamma\gamma,~b\bar{b}\tau^+\tau^-,~b\bar{b}b\bar{b},~b\bar{b}VV$~($V=W,Z$) search results~\cite{Sirunyan:2018ayu}. Part of this region (at very small $\tb \sim 1$) is furthermore constrained by charged Higgs boson searches in the $pp\to tbH^\pm \to tb(tb)$ channel~\cite{Aaboud:2018cwk}.

The \emph{hatched region} of \fig{fig:mh125-ltb_msusy} is disfavored at the $2\sigma$ level by the \sm{}-like Higgs-boson rate measurements at the \lhc{}, as evaluated by \HiSi. Its boundary is located at around $M_A\simeq 650$\,GeV and depends only mildly on $\tb$. Lower $M_A$ values are excluded as the coupling of the light Higgs boson to bottom quarks is enhanced with respect to the \sm{} prediction. At larger $M_A$ values this coupling approaches the \sm{} value, as expected in the decoupling limit. At very small $\tb$, the Higgs rate measurements exclude values up to $M_A \simeq 700$\,GeV due to a suppression of the light Higgs-boson gluon-fusion cross section by a few percent. This originates from a slight suppression of the light Higgs-boson coupling to top quarks induced by higher-order corrections to the external Higgs leg (accounted for by employing the $\mathbf{Z}$-matrix).

The \emph{green contours} in \fig{fig:mh125-ltb_msusy} indicate the \msusy values required to reach $M_h\simeq 125$\,GeV. While only moderate \msusy values of up to $10^4$\,GeV are required for $\tb\gtrsim 8$, much higher values up to around $10^8$\,GeV are needed for $\tb\gtrsim 1.5$. Moreover, if $M_A$ is below 500~GeV, $\msusy$ has to be raised to very large values of $10^{10}$\,GeV and higher. As mentioned above, for very low $M_A\lesssim 200$\,GeV, \msusy would have to be raised above $10^{16}$\,GeV to obtain $M_h \ge 122$\,GeV, a region that we disregard in our work.

\begin{figure}[t]
\begin{center}
\vspace*{-4mm}
\includegraphics[width=0.75\textwidth]{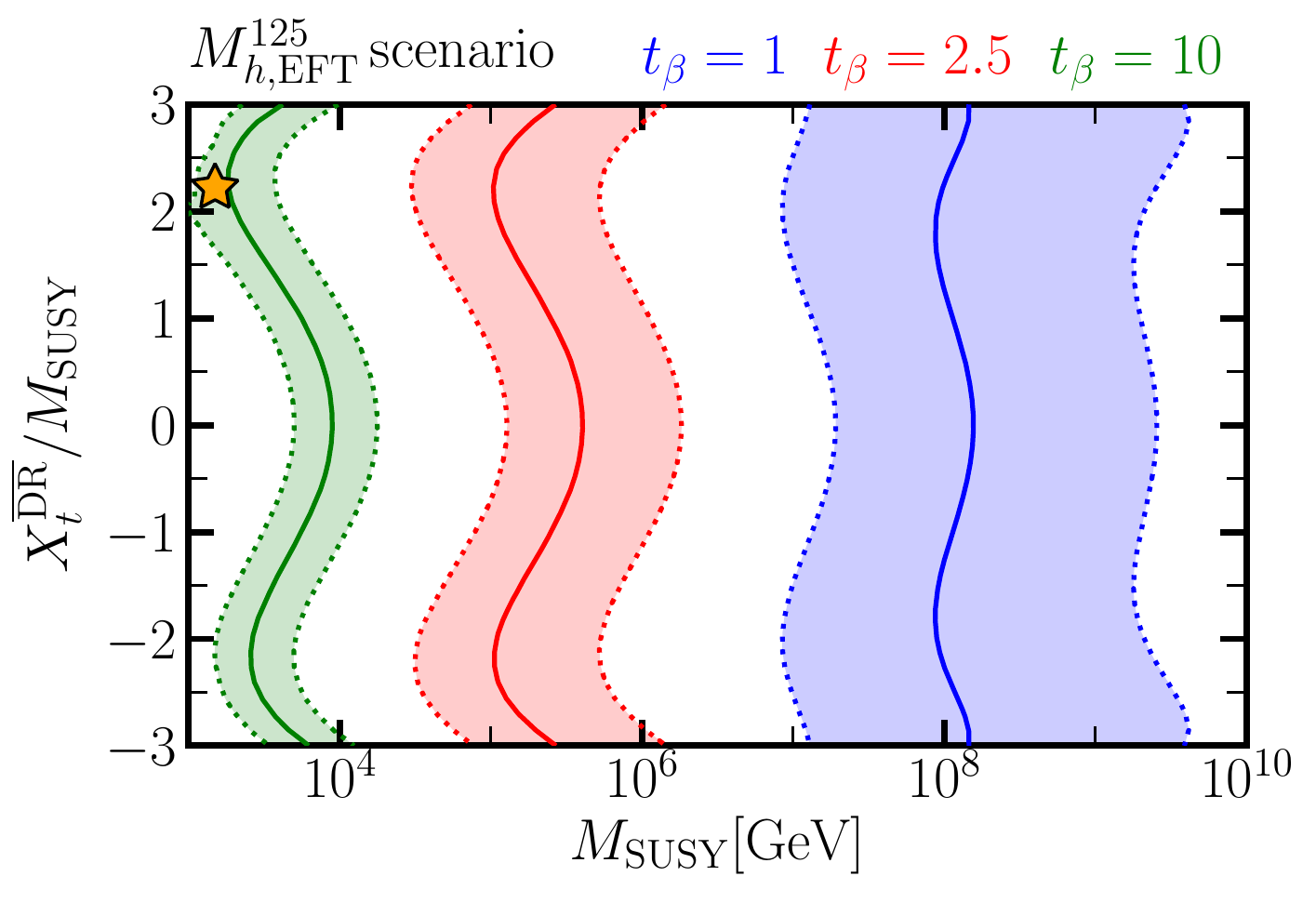}
\end{center}
\vspace*{-10mm}
\caption{Contour lines of $M_h\simeq 125$\,GeV (\emph{solid}) and $M_h\simeq 122,128$\,GeV (\emph{dashed}) as a function of $\msusy$ and $X_t^\text{\drbar}/\msusy$, for $\tb=1$ (\emph{blue}), $\tb=2.5$ (\emph{red}) and $\tb=10$ (\emph{green}).  $M_A$ is fixed to 1~TeV. All remaining parameters are as in the \ltbmhsc scenario. The \emph{orange star} marks the approximate position of the \mhsc scenario~\cite{Bahl:2018zmf}.}
\label{fig:mh125-ltb_msusyxt}
\end{figure}

We further explore the required \susy{} scale in \fig{fig:mh125-ltb_msusyxt}, which shows $M_h = 125$\,GeV contours (\emph{solid}) as a function of $\msusy$ and $X_t^\text{\drbar}/\msusy$ (with $X_t = A_t - \mu/\tb$) for $\tb=1$ (\emph{blue}), $\tb=2.5$ (\emph{red}) and $\tb=10$ (\emph{green}) and $M_A=1\,$TeV. Note that in the \ltbmhsc benchmark scenario, as defined in \eqns{eq:commonpara}{eq:ltbmhscscenario}, $X_t^\text{\drbar}/\msusy$ is almost zero. Clearly, the needed \msusy value strongly depends on the chosen $\tb$ value. While for $\tb = 10$ values of $\mathcal{O}(10^3\,\GeV)$ are sufficient, for $\tb = 1$ values of $\mathcal{O}(10^8\,\GeV)$ are needed. We also observe that for high \susy{} scales there is only a mild dependence on $X_t^\text{\drbar}/\msusy$. This behavior reflects the decrease of the strong gauge coupling and the top-quark Yukawa coupling, which multiply the dominant threshold corrections between the full \mssm{} and the \eft{} below the \susy{} scale involving $X_t^\text{\drbar}/\msusy$, with rising \msusy. Therefore, at low $\tb$ values, choosing a high value of $|X_t^\text{\drbar}/\msusy|$ does not allow to significantly lower the required \susy{} scale. The \emph{dashed lines} mark the contours of $M_h = 122$\,GeV and $M_h = 128$\,GeV, indicating the allowed region when taking into account a simple global theoretical uncertainty estimate of $3$\,GeV on the Higgs mass calculation \cite{Degrassi:2002fi,Allanach:2004rh}. The broadening of the corresponding colored band with rising \msusy corresponds to a growing uncertainty in the deduction of the required \susy{} scale.

For comparison, we also show in \fig{fig:mh125-ltb_msusyxt} the approximate position of the \mhsc scenario of Ref.~\cite{Bahl:2018zmf} in the considered parameter plane (\emph{orange star}). In the \mhsc scenario the masses of the third-generation squarks is set to $1.5$\,TeV. A large stop mixing parameter of $X_t^{\text{OS}} = 2.8$\,GeV is needed to obtain $M_h\simeq 125$\,GeV.\footnote{In the \mhsc scenario, the OS scheme is used for the renormalization of the stop parameters. We converted the OS parameters at the one-loop level to the \drbar\ scheme to obtain the approximate position in the $(\msusy,\, X_t^\text{\drbar}/\msusy)$ plane shown in \fig{fig:mh125-ltb_msusyxt}.}

\begin{figure}[t]
\begin{center}
\includegraphics[width=0.49\textwidth]{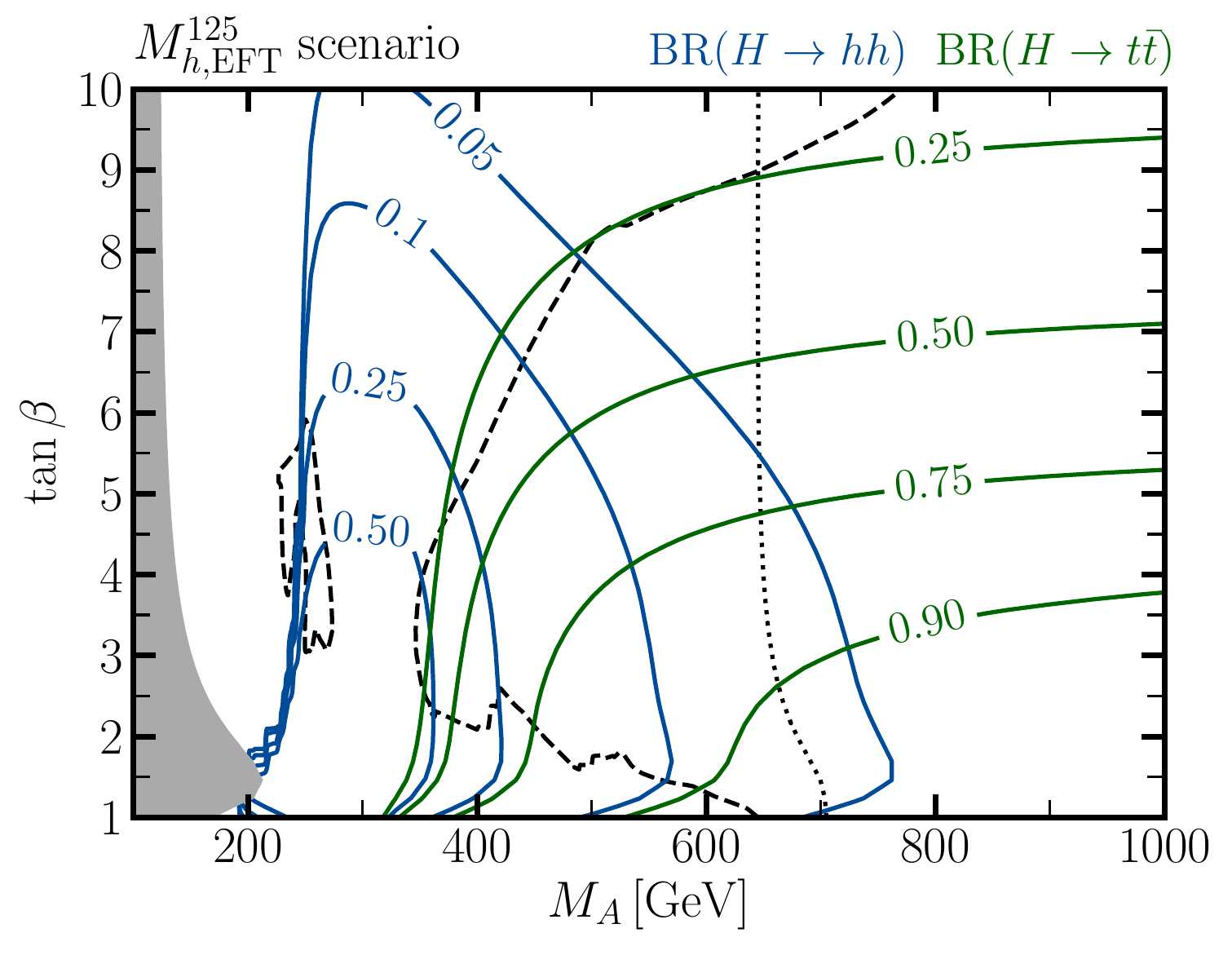}\hfill
\includegraphics[width=0.49\textwidth]{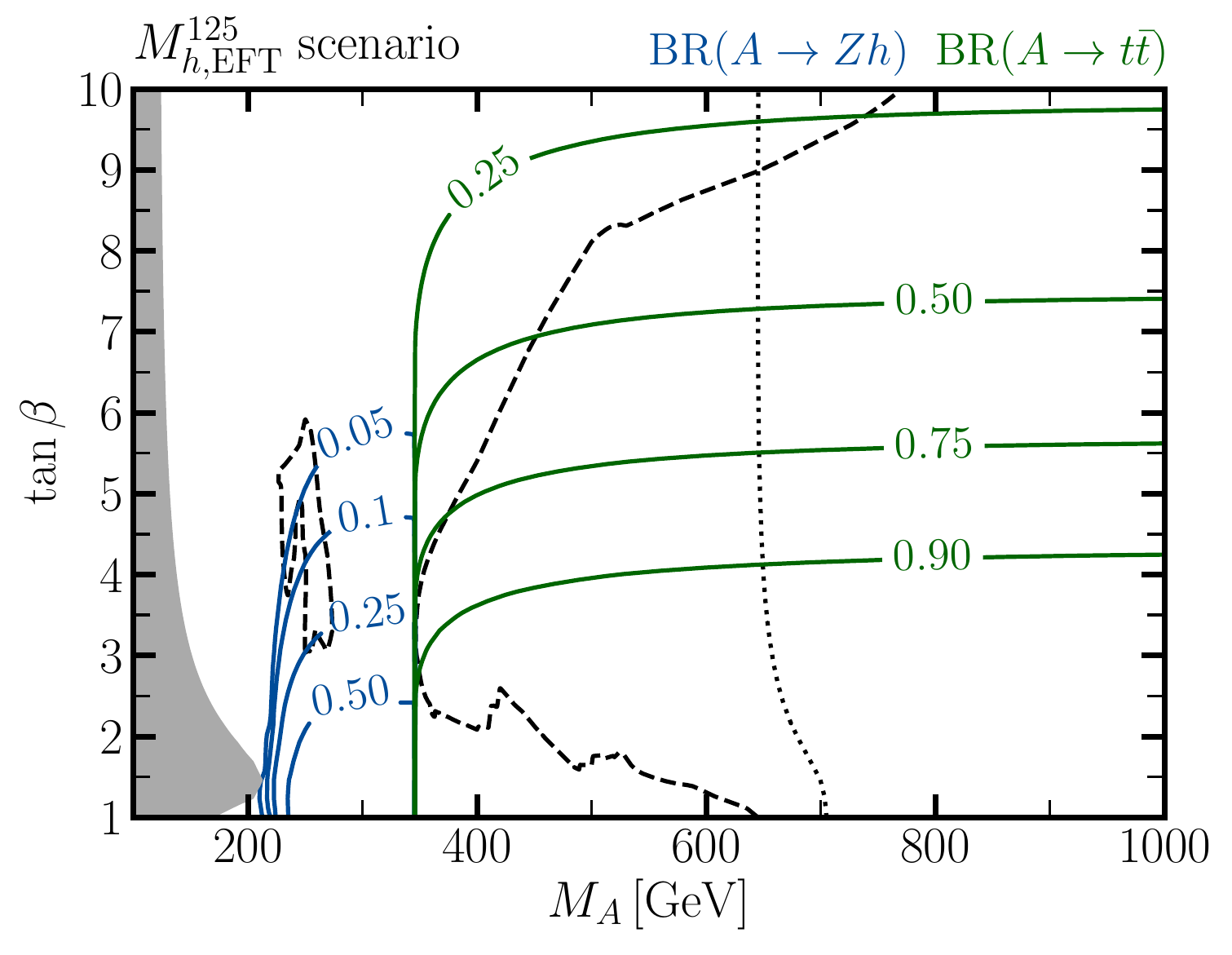}
\end{center}
\vspace*{-10mm}
\caption{\emph{Left}: Branching ratios of the heavy \cp-even Higgs boson $H$ into a pair of the light \cp-even Higgs bosons $h$ (\emph{blue}) and into a pair of top quarks (\emph{green}) as a function of $\mA$ and $\tb$ in the \ltblchi\ scenario. \emph{Right}: Branching ratios of the \cp-odd Higgs boson $A$ into a $Z$-boson and the light \cp-even Higgs boson $h$ (\emph{blue}) and into a pair of top quarks (\emph{green}). In each plot, the gray exclusion region and the boundaries of the blue and the hatched exclusion regions (shown as \emph{dashed} and \emph{dotted black lines}, respectively) of \fig{fig:mh125-ltb_msusy} are also depicted.}
\label{fig:mh125-ltb_BRAH}
\end{figure}

We explore the decays of the \cp{}-even $H$ and \cp{}-odd $A$~bosons in \fig{fig:mh125-ltb_BRAH} in the parameter region of $M_A \le 1\,\TeV$. The \emph{left panel} shows the branching ratios of the $H$~boson decays into a pair of light Higgs bosons, $H\to hh$ (\emph{blue contours}), and into a pair of top quarks, $H\to t\bar{t}$ (\emph{green contours}). At low values $M_H \lesssim 2 M_t$, the decay $H\to hh$ can reach a branching ratio of more than $50\%$ (at low $\tb$). It mostly competes with the decays $H\to VV$ ($V=W,Z$) in this mass regime.  Once $M_A$ is raised above the kinematic threshold for the $H\to t\bar{t}$ decay, $M_H\gtrsim 2 M_t$, the decay $H\to t\bar{t}$ becomes the dominant mode and suppresses the $H\to hh$ and $H\to VV$ decays. In the unexcluded region ($M_A\gtrsim 650$\,GeV), the branching ratio for $H\to hh$ drops below 10\%, while the decay $H\to t\bar{t}$ can reach values above $90\%$ for low $\tb$. If $\tb$ is increased the $H$ boson coupling to top quarks becomes suppressed while its coupling to bottom quarks and tau leptons becomes enhanced, such that the decay $H\to b\bar{b}$ eventually becomes dominant. The \emph{right panel} of \fig{fig:mh125-ltb_BRAH} shows the branching ratios of the $A$~boson decays into a $Z$ boson and a $h$~boson, $A\to Zh$ (\emph{blue contours}), and into a top-quark pair, $A\to t\bar{t}$ (\emph{green contours}). We observe a qualitatively similar behavior as for the $H$ decays. However, below the top-quark pair production threshold, the $A\to Zh$ decay competes against the decay into a bottom-quark pair, $A\to b\bar{b}$, which is dominant for $\tb\gtrsim 3$. Beyond its kinematic threshold, $M_A \simeq 2 M_t$, the decay $A\to t\bar{t}$ quickly becomes dominant, and the decay $A\to Zh$ is negligible. In this mass regime we therefore expect that upcoming dedicated \lhc{} searches for heavy Higgs bosons decaying to top-quark pairs, see \citere{Aaboud:2017hnm} for a Run-1 analysis by ATLAS, will be an excellent probe. Thus, we encourage the experiments to perform such an analysis.
Interference effects between the signal $gg\to H/A\to t\bar t$ and the background $gg\to t\bar t$ are known to be large, see \citeres{Gaemers:1984sj,Dicus:1994bm,Moretti:2012mq,Frederix:2007gi,Hespel:2016qaf,Djouadi:2019cbm},
and due to the heavy \susy{} spectrum in our example can be parametrized as a function of the Higgs masses and $\tb$ for $M_A$ values in the decoupling limit ($M_A\gg M_Z$).


\subsection{\texorpdfstring{$\boldsymbol{M}_{\boldsymbol{h},\text{\eft{}}}^{\text{125}}\boldsymbol{(\tilde\chi)}$}{MhEFT125(chi)} scenario}
The second benchmark scenario we propose is the \ltblchi scenario. In contrast to the \ltbmhsc scenario, this scenario features light neutralinos and charginos whose presence significantly alters the Higgs phenomenology. We choose for the Higgsino, bino and wino mass parameters
\begin{eqnarray}
&\mu=180~\text{GeV},\,\quad
M_1=160~\text{GeV},\quad M_2=180~\text{GeV}\,,
\label{eq:ltblchiscenario}
\end{eqnarray}
respectively, such that this scenario represents an extension of the \lchi scenario~\cite{Bahl:2018zmf} to low $\tb$ values. The other input parameters are fixed according to \eqn{eq:commonpara}, and the \susy{} scale is again adjusted at every parameter point in order to obtain a light Higgs mass of $M_h\simeq 125$\,GeV.

\begin{figure}[t]
\begin{center}
\vspace*{-4mm}
\includegraphics[width=0.75\textwidth]{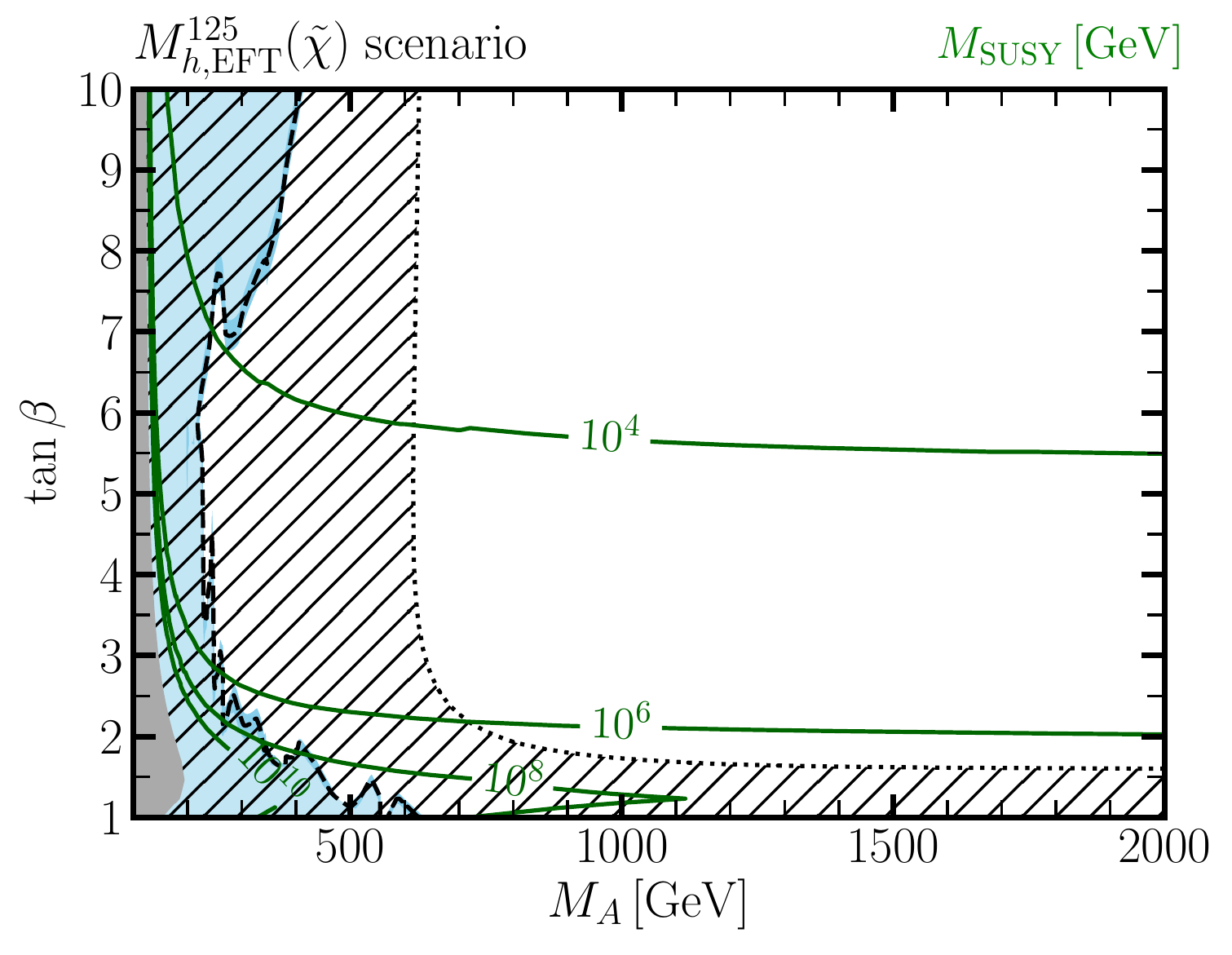}
\end{center}
\vspace*{-10mm}
\caption{The \ltblchi benchmark scenario shown in the $(\mA\,,\tb)$ plane. The \emph{green solid contour lines} show the required \susy{} scale. The \emph{black dotted line} and \emph{hatched area} depict the $2\sigma$ disfavored region arising from the Higgs boson signal rate measurements. The \emph{blue region} with the \emph{black dashed boundary} is excluded at the $95\%~\mathrm{C.L}$ by \lhc{} searches for additional Higgs bosons (the \emph{dark blue band} shows how the theoretical rate uncertainty affects the exclusion). The \emph{gray area} is excluded because the mass of the \sm{}-like Higgs boson $h$ is below $122$\,GeV.}
\label{fig:mh125-ltblchi_msusy}
\end{figure}

In \fig{fig:mh125-ltblchi_msusy} we present the \ltblchi scenario in the ($M_A$, $\tb$) parameter plane. The \emph{green contour lines} show again the \msusy values required to reach $M_h\simeq 125$\,GeV. The presence of light electroweakinos leads to an upwards shift of the \sm{}-like Higgs-boson mass of $\sim 1.5$\,GeV, therefore, smaller \msusy values are required as compared to the previous scenario. Again, we encounter a parameter region (\emph{gray area}) at very low $M_A\lesssim 200$\,GeV, for which \msusy would have to be chosen above $10^{16}$\,GeV to obtain $M_h\ge 122$\,GeV. However, in comparison to the \ltbmhsc scenario, this region is slightly smaller due to the aforementioned contribution of the light electroweakinos to the light Higgs-boson mass.

We again show in \fig{fig:mh125-ltblchi_msusy} the direct constraints from \lhc{} searches for non-\sm{}-like Higgs bosons (\emph{dashed black line} and \emph{blue area}, with the \emph{dark blue band} indicating the theoretical uncertainty) and indirect constraints from \sm{}-like Higgs-boson signal-rate measurements (\emph{dotted black line} and \emph{hatched area}). The excluded area arising from direct searches for non-\sm{}-like Higgs bosons --- the relevant search channels are the same as in the \ltbmhsc scenario (see above) --- is smaller than in the previous case. In particular, $pp\to H/A\to \tau^+\tau^-$ searches only exclude $M_A$ values up to $220~(400)$\,GeV for moderately large $\tb$ values of $6$ ($10$). For $\tb$ values between $2$ to $6$ the parameter space for $M_A \lesssim 250$\,GeV is constrained mostly by $pp \to H\to ZZ$ searches. The low $\tb$ region is again constrained by $pp\to H\to hh$ searches and charged Higgs-boson searches. The decrease in sensitivity of all these search channels with respect to the previous scenario arises from the fact that the heavy Higgs bosons $H$ and $A$ can decay to pairs of light electroweakinos. If such decays have sizable rates, the branching ratios for heavy Higgs-boson decays to \sm{} particles are suppressed. We will discuss the heavy Higgs-boson decays to electroweakinos in detail below.

In \fig{fig:mh125-ltblchi_msusy} the parameter region at $M_A\lesssim 630$\,GeV is excluded by the \sm{}-like Higgs-boson signal-strength measurements. In comparison to the \ltbmhsc scenario, however, an additional exclusion arises in the region of very small $\tb$ values $\lesssim 1.5$ (depending on $M_A$). In this area the branching ratio of the light Higgs decay to two photons, $h\to \gamma\gamma$, is significantly enhanced due to the presence of light charginos. We will also explore this decay mode below.

\begin{figure}[t]
\begin{center}
\vspace*{-4mm}
\includegraphics[width=0.75\textwidth]{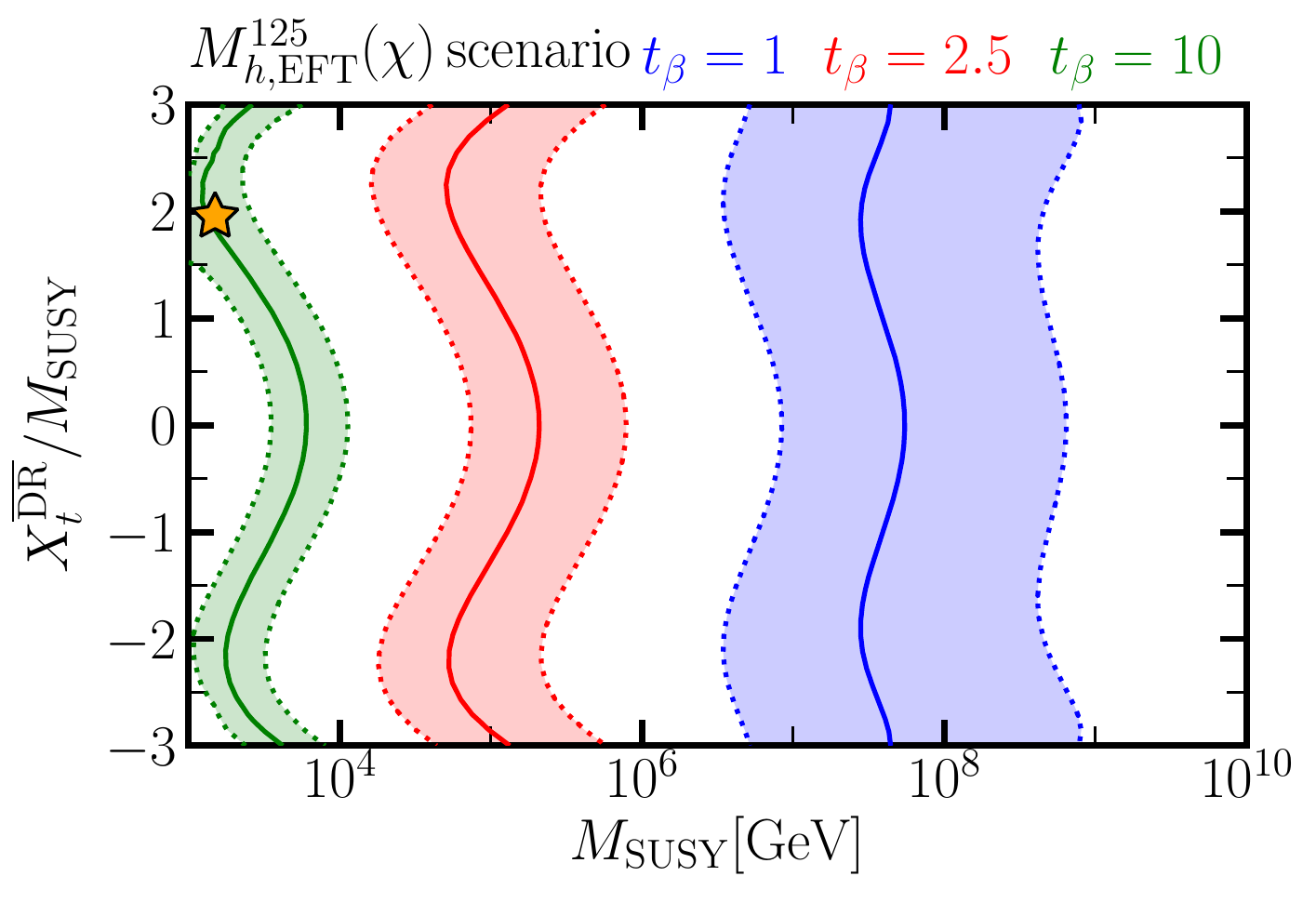}
\end{center}
\vspace*{-10mm}
\caption{Contour lines of $M_h\simeq 125$\,GeV (\emph{solid}) and $M_h\simeq 122,128$\,GeV (\emph{dashed}) as a function of $\msusy$ and $X_t^\text{\drbar}/\msusy)$, for $\tb=1$ (\emph{blue}), $\tb=2.5$ (\emph{red}) and $\tb=10$ (\emph{green}). $M_A$ is fixed to $1$\,TeV. All remaining parameters are as in the \ltblchi scenario. The \emph{orange star} marks the approximate position of the \lchi scenario~\cite{Bahl:2018zmf}.}
\label{fig:mh125-ltbchi_msusyxt}
\end{figure}

\fig{fig:mh125-ltbchi_msusyxt} shows the $M_h = 125$\,GeV contours (\emph{solid}) around the \ltblchi scenario as a function of $\msusy$ and $X_t^\text{\drbar}/\msusy$ for $\tb=1$ (\emph{blue}), $\tb=2.5$ (\emph{red}) and $\tb=10$ (\emph{green}). In comparison to the previous benchmark scenario, see \fig{fig:mh125-ltb_msusyxt}, the Higgs-boson mass contours are shifted by approximately half an order of magnitude to lower \msusy values. As mentioned before, this shift originates from the presence of light electroweakinos in this scenario which lead to an upwards shift of the light Higgs-boson mass. For comparison, we also show the approximate position of the $M_h^{125}(\tilde\chi)$ scenario of \citere{Bahl:2018zmf}, for which $\msusy = 1.5$\,TeV and $X_t^{\text{OS}} = 2.5$\,TeV. Keep in mind the different renormalization schemes: \citere{Bahl:2018zmf} employs $X_t^\text{\text{OS}}$, whereas this work is based on $X_t^\text{\drbar}$.

\begin{figure}[t]
\begin{center}
\includegraphics[width=0.49\textwidth]{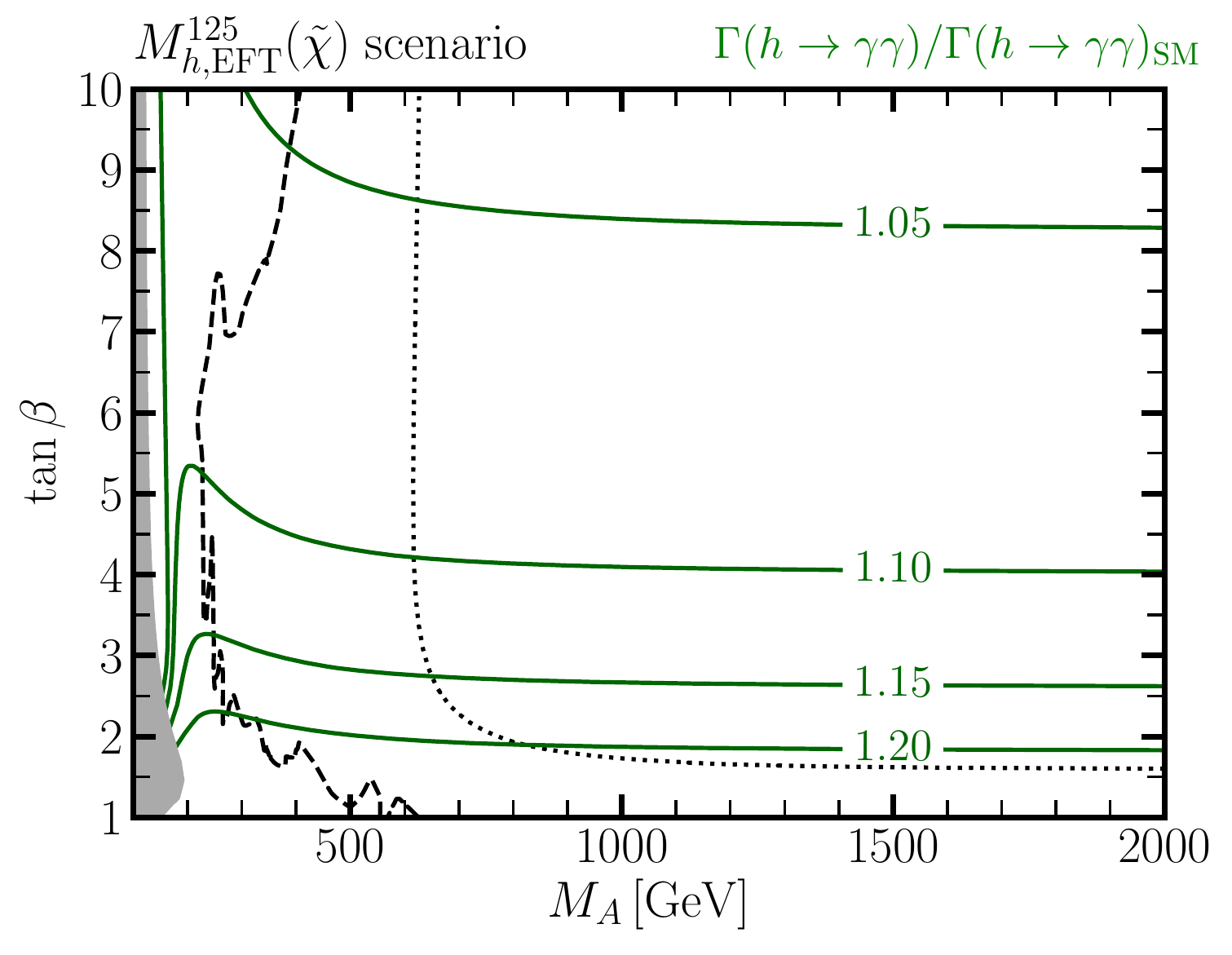}\hfill
\includegraphics[width=0.49\textwidth]{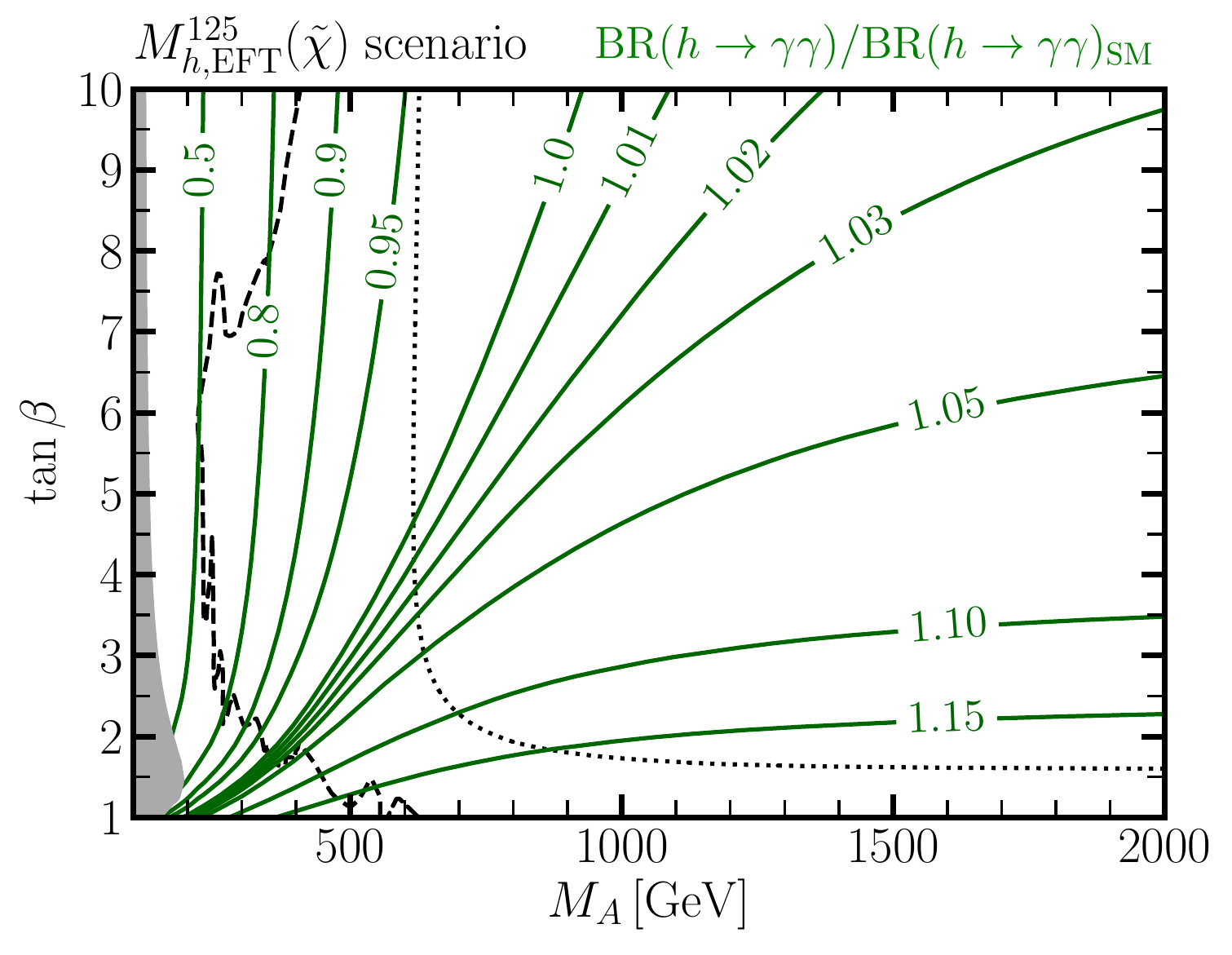}
\end{center}
\vspace*{-10mm}
\caption{\emph{Left}: Decay width of the light \cp-even Higgs boson $h$ into photons (\emph{green contour lines}) in the ($\mA$, $\tb$) plane in the \ltblchi scenario, normalized to the corresponding \sm{} prediction. \emph{Right}: Branching ratio of the decay $h\to \gamma \gamma$, normalized to its \sm{} prediction (\emph{green contour lines}). In each plot, the gray exclusion region and the boundaries of the blue and the hatched exclusion regions (shown as \emph{dashed} and \emph{dotted black lines}, respectively) of \fig{fig:mh125-ltblchi_msusy} are also depicted.}
\label{fig:mh125-ltblchi_hgaga}
\end{figure}

We now discuss in detail the impact of the light electroweakinos on the decays of the \mssm{} Higgs bosons.
Such decays were already considered in early discussions of discovery prospects of the CMS detector~\cite{Denegri:2001pn,Moortgat:2001pp,Abdullin:2005yn}
and were advocated by theorists in recent years~\cite{Arganda:2012qp,Arganda:2013ve,Craig:2015jba,Ananthanarayan:2015fwa,Barman:2016kgt,Barman:2016jov,Arganda:2017wjh,Profumo:2017ntc,Aboubrahim:2018tpf,Bahl:2018jom},
see e.g.~\citere{Gori:2018pmk} for a thorough analysis on the sensitivity in a class of benchmark scenarios.
The electroweakino spectrum is fixed at tree-level by the choice of $\mu$, $M_1$, $M_2$ and the value of $\tb$ and exhibits a strong wino-Higgsino mixing in both the neutralino and chargino sector. This mixing,
which pushes the wino and Higgsino mass eigenstates away from each other, is enhanced for small values of $\tb$, such that at $\tb=1$ the spectrum is slightly less compressed than at $\tb=10$.
The lightest neutralino mass increases from $\sim 85$\,GeV to $\sim 112$\,GeV between $\tb=1$ and $\tb=10$.
In the \emph{left panel} of \fig{fig:mh125-ltblchi_hgaga} we show the partial width of the decay $h\to\gamma\gamma$ normalized to the \sm{} prediction. The decay width is enhanced  by $\gtrsim 20\%$ for low $\tb \lesssim 2$. This enhancement originates from loop corrections involving light charginos. As discussed above, our choice of $M_2 = \mu$ in this scenario leads to a significant wino-Higgsino mixing in the chargino sector, which, in turn, results in a large coupling of the charginos to the \mssm{} Higgs bosons. Hence, we have a sizable  contribution to $h\to \gamma\gamma$ from charginos in this scenario. The distortions for $M_A\lesssim 200$\,GeV are due to sizable mixing effects between the $h$ and $H$ bosons. In the \emph{right panel} of \fig{fig:mh125-ltblchi_hgaga} we show the branching ratio of the decay $h\to\gamma\gamma$ normalized to its \sm{} prediction. In comparison to the partial width, the enhancement of the $h\to b\bar b$ decay width in the low-$M_A$ regime leads to an additional suppression of the branching ratio of the $h\to \gamma\gamma$ decay. For large $M_A$ values above around $800$\,GeV and $\tb\lesssim 2$, the branching ratio is enhanced by $\gtrsim 15\%$, which yields the exclusion from the $h\to\gamma\gamma$ signal rate measurement at very low $\tb$ values, which persists in the Higgs decoupling regime ($M_A \gg M_Z$). Future precision measurements of this Higgs-boson decay mode thus offer the possibility to indirectly probe for light electroweakinos within this scenario, even if the remaining Higgs bosons are very heavy. On the other hand, if the branching ratio $h\to \gamma\gamma$ remains to be consistent with the \sm{} predictions in the future, the lower bound on $\tb$ will be increased. This will in turn lead to a more stringent upper bound on the \susy{} scale (see \fig{fig:mh125-ltbchi_msusyxt}).

\begin{figure}[t]
\begin{center}
\includegraphics[width=0.49\textwidth]{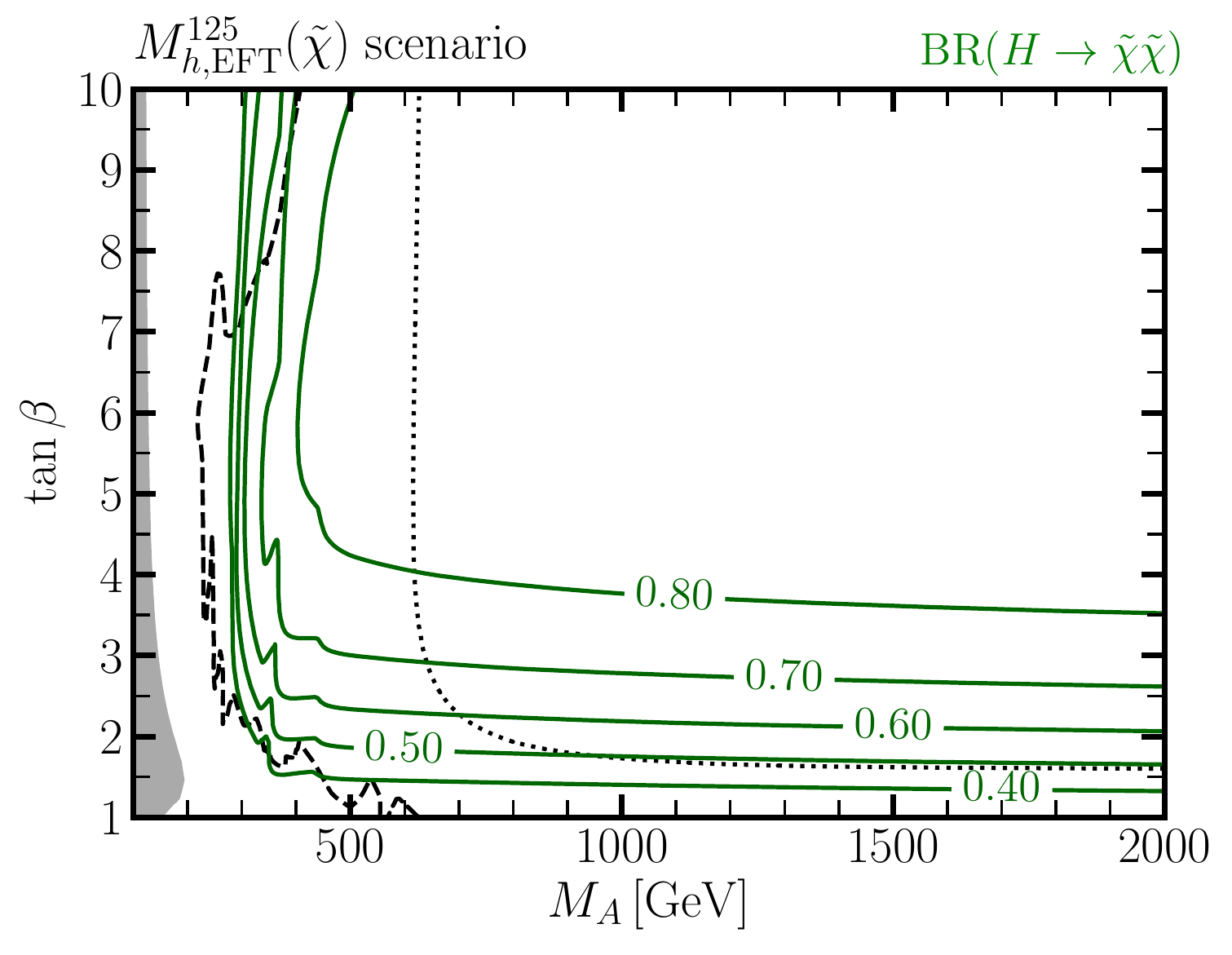}\hfill
\includegraphics[width=0.49\textwidth]{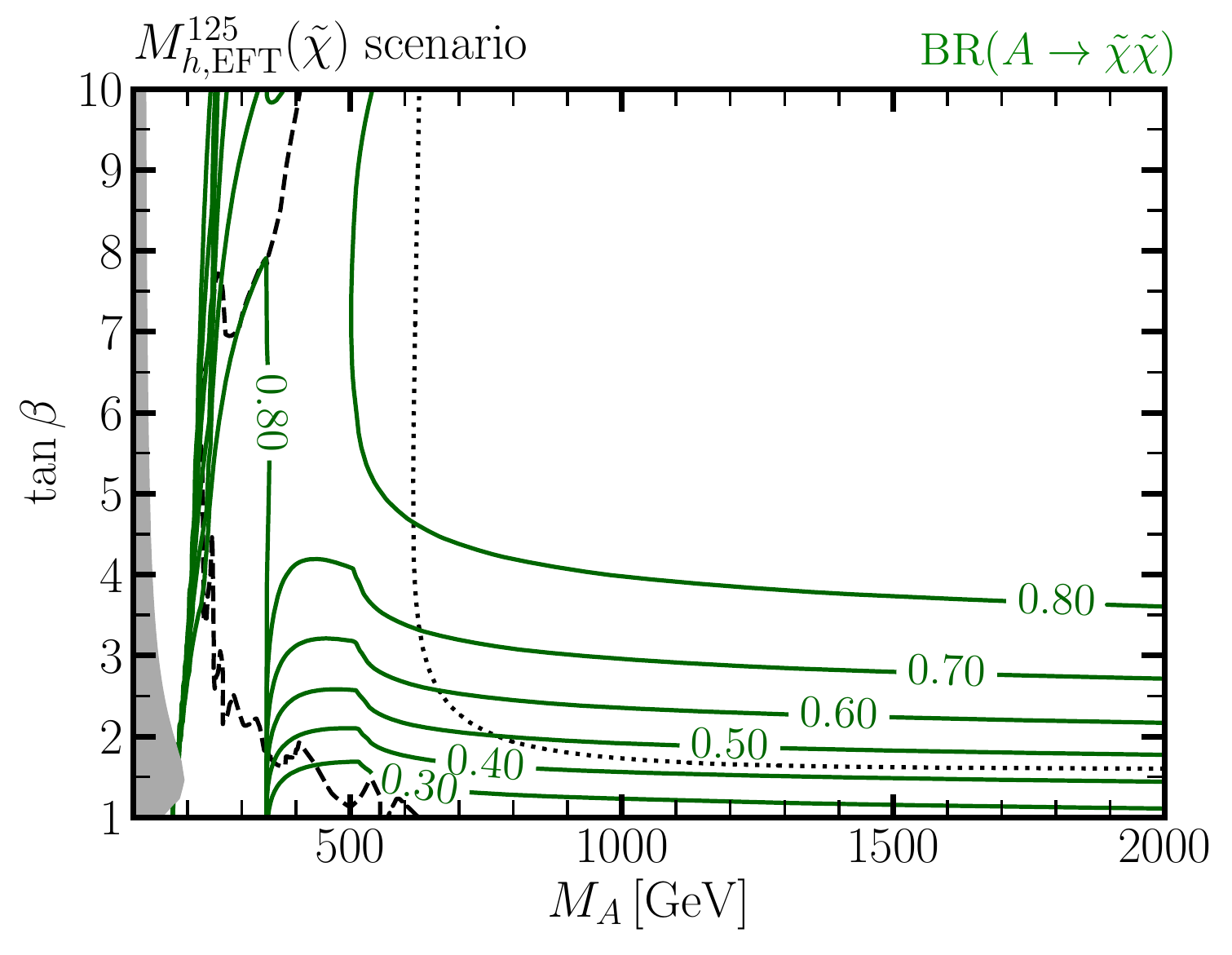}
\end{center}
\vspace*{-10mm}
\caption{Branching ratios of the decays of the heavy \cp-even Higgs boson $H$ (\emph{left}) and the \cp-odd Higgs boson $A$ (\emph{right}) into electroweakino pairs, shown in the ($\mA$, $\tb$) plane  in the \ltblchi scenario. The contributions from all kinematically allowed combinations of electroweakinos in the final state are summed. In each plot, the gray exclusion region and the boundaries of the blue and the hatched exclusion regions (shown as \emph{dashed} and \emph{dotted black lines}, respectively) of \fig{fig:mh125-ltblchi_msusy} are also depicted.}
\label{fig:mh125-ltblchi_HAtoSUSY}
\end{figure}

\fig{fig:mh125-ltblchi_HAtoSUSY} shows the branching ratios for the decays of the heavy Higgs bosons $H$ (\emph{left panel}) and $A$ (\emph{right panel}) into pairs of charginos and neutralinos. The contributions from all kinematically accessible electroweakino final states are summed. For both $H$ and $A$ the branching ratio into electroweakinos exceeds $80\%$ for $\tb\gtrsim 4$ and $M_A\gtrsim 500$\,GeV. When decreasing $M_A$ below $500$\,GeV, we encounter kinematic thresholds where some decay modes into electroweakinos become inaccessible, leading to a gradual decrease with sharp transitions of the inclusive Higgs-to-electroweakino  branching ratio. The structures at $M_A\sim 340$\,GeV are caused by the kinematic threshold for the decays into a pair of top quarks. The large branching ratios for the decays into electroweakinos strongly motivates dedicated \lhc{} searches for these signatures. We will discuss the most promising signatures in detail below.

\begin{figure}[t]
\begin{center}
\includegraphics[width=0.49\textwidth]{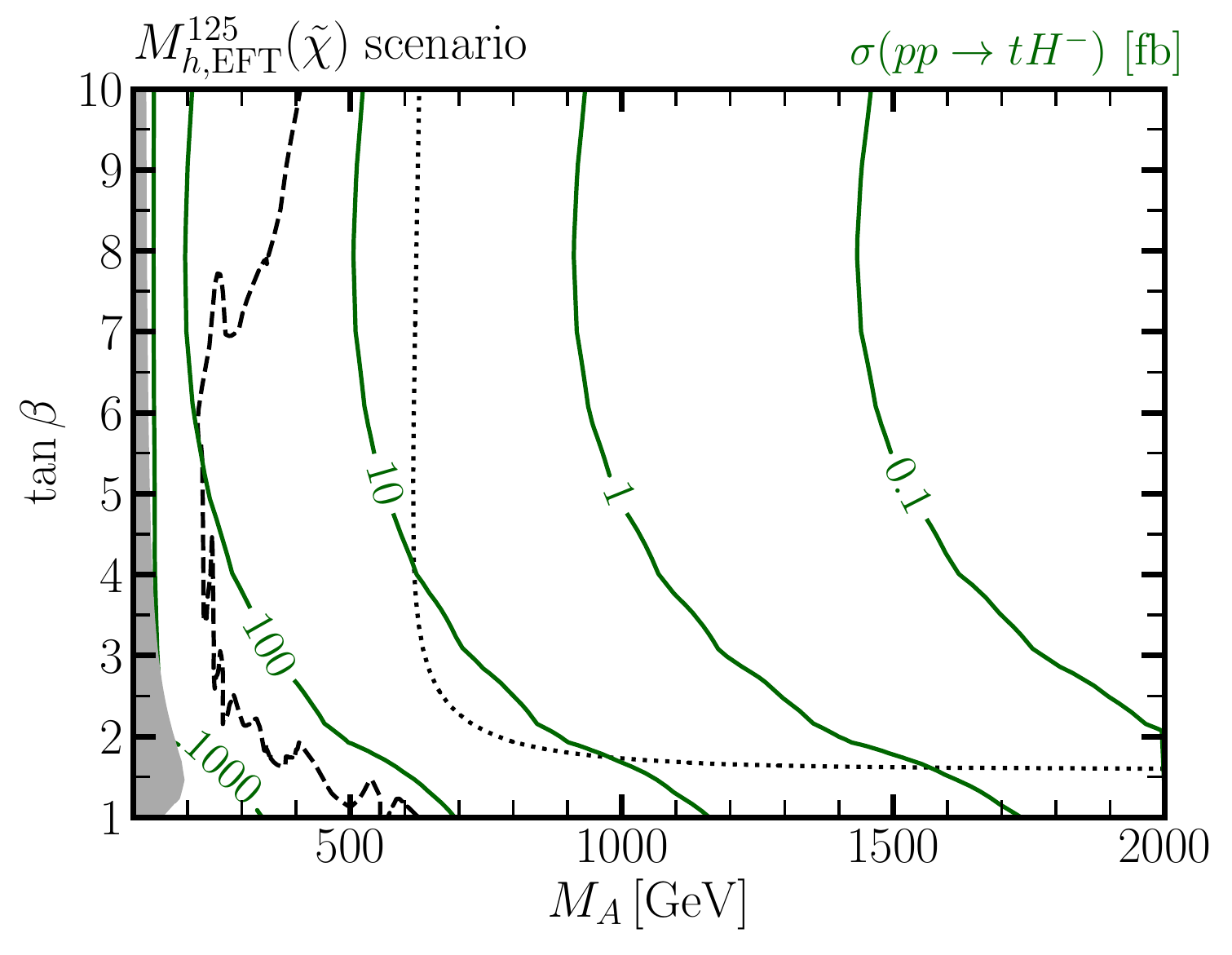}\hfill
\includegraphics[width=0.49\textwidth]{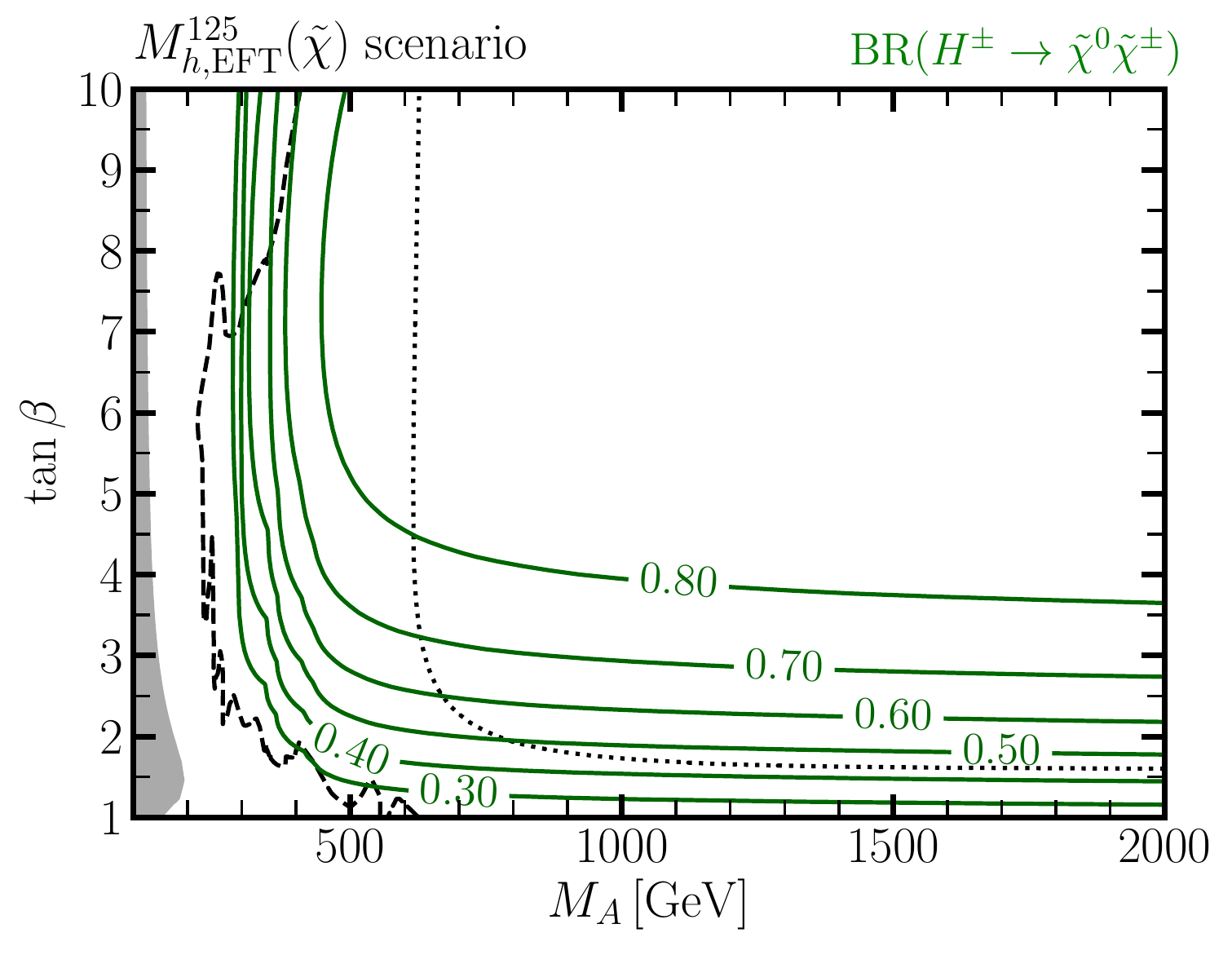}
\end{center}
\vspace*{-10mm}
\caption{Charged Higgs-boson phenomenology in the \ltblchi scenario. \emph{Left}: Production cross section (in fb) of a negatively charged Higgs boson $H^-$ in association with a top quark (\emph{green contour lines}) at the \lhc{} with $13\,\mathrm{TeV}$ center-of-mass energy. \emph{Right}: Branching ratio of the charged Higgs boson decaying into chargino--neutralino pairs (\emph{green contour lines}). The contributions from all kinematically allowed combinations of electroweakinos in the final state are summed. In each plot, the gray exclusion region and the boundaries of the blue and the hatched exclusion regions (shown as \emph{dashed} and \emph{dotted black lines}, respectively) of \fig{fig:mh125-ltblchi_msusy} are also depicted.}
\label{fig:mh125-ltblchi_Hpm}
\end{figure}

The presence of light electroweakinos also affects the decay rates of the charged Higgs boson $H^\pm$. In the \emph{right panel} of \fig{fig:mh125-ltblchi_Hpm} we show the branching ratio of the charged Higgs boson decaying into neutralino-chargino pairs. Again, the contributions from all kinematically accessible electroweakino final states are summed. Similar as for the neutral Higgs bosons, the branching ratio for charged Higgs-boson decays to electroweakinos exceeds $80\%$ for large $M_A \gtrsim 700$\,GeV and $\tb \gtrsim 4$. We furthermore provide in the \emph{left panel} of \fig{fig:mh125-ltblchi_Hpm} the production cross section (in fb) for top-quark associated production of a negatively charged Higgs boson (the charge-conjugated process has identical production rate) at the \lhc{} with a center-of-mass energy of $13$\,TeV. While most of the parameter region with large cross section is already strongly constrained by the \sm{}-like Higgs-boson signal-rate measurements, the production cross section can still exceed $10$\,fb in the allowed parameter space at $\tb\sim 3$ and $M_A\sim 700$\,GeV. Together with a decay rate to electroweakinos of $60$--$70\%$, the signal cross section for $pp\to t H^\pm \to t (\chi^\pm \chi)$ can still be $\gtrsim 14~\mathrm{fb}$, corresponding to more than $2200$ expected signal events in the currently recorded integrated luminosity of \lhc{} Run-2 per experiment, $\mathcal{L}_\text{int} \simeq 160~\mathrm{fb}^{-1}$. Thus, searches for a charged Higgs boson decaying into electroweakinos could be a promising way to further probe the \ltblchi scenario.

\begin{figure}[t]
\begin{minipage}{.5\textwidth}
\includegraphics[width=\textwidth]{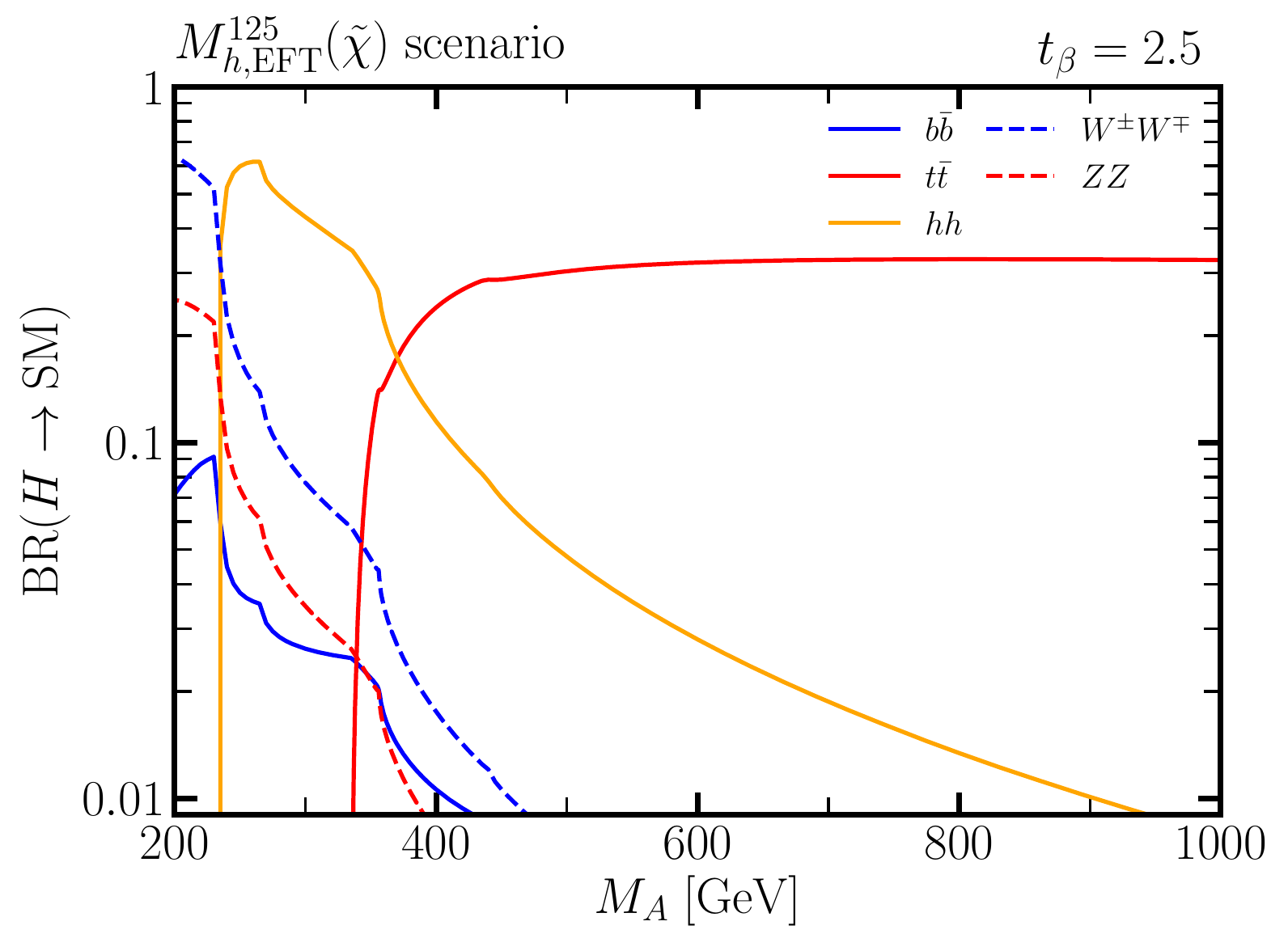}
\end{minipage}
\begin{minipage}{.5\textwidth}
\includegraphics[width=\textwidth]{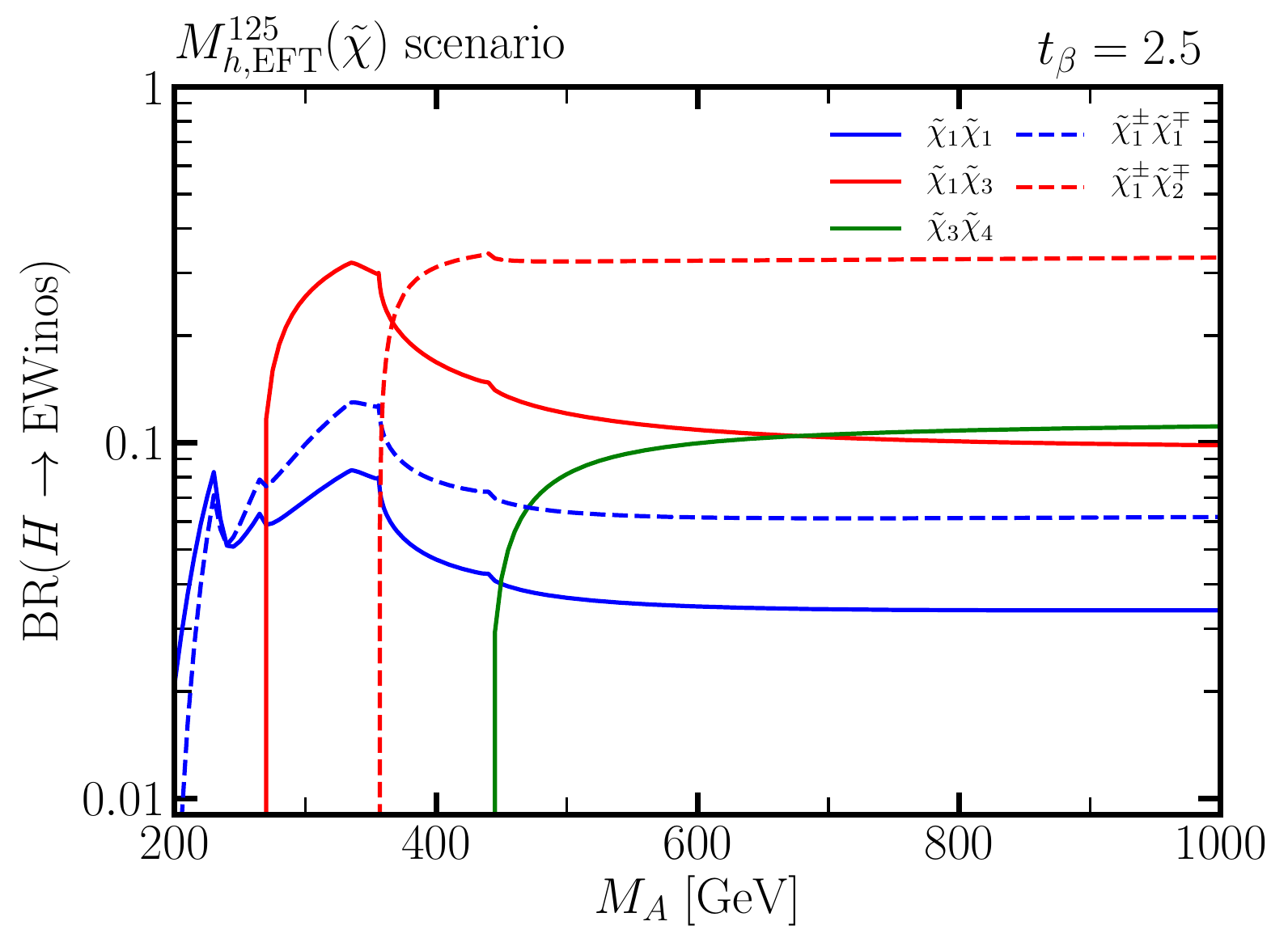}
\end{minipage}
\begin{minipage}{.5\textwidth}
\includegraphics[width=\textwidth]{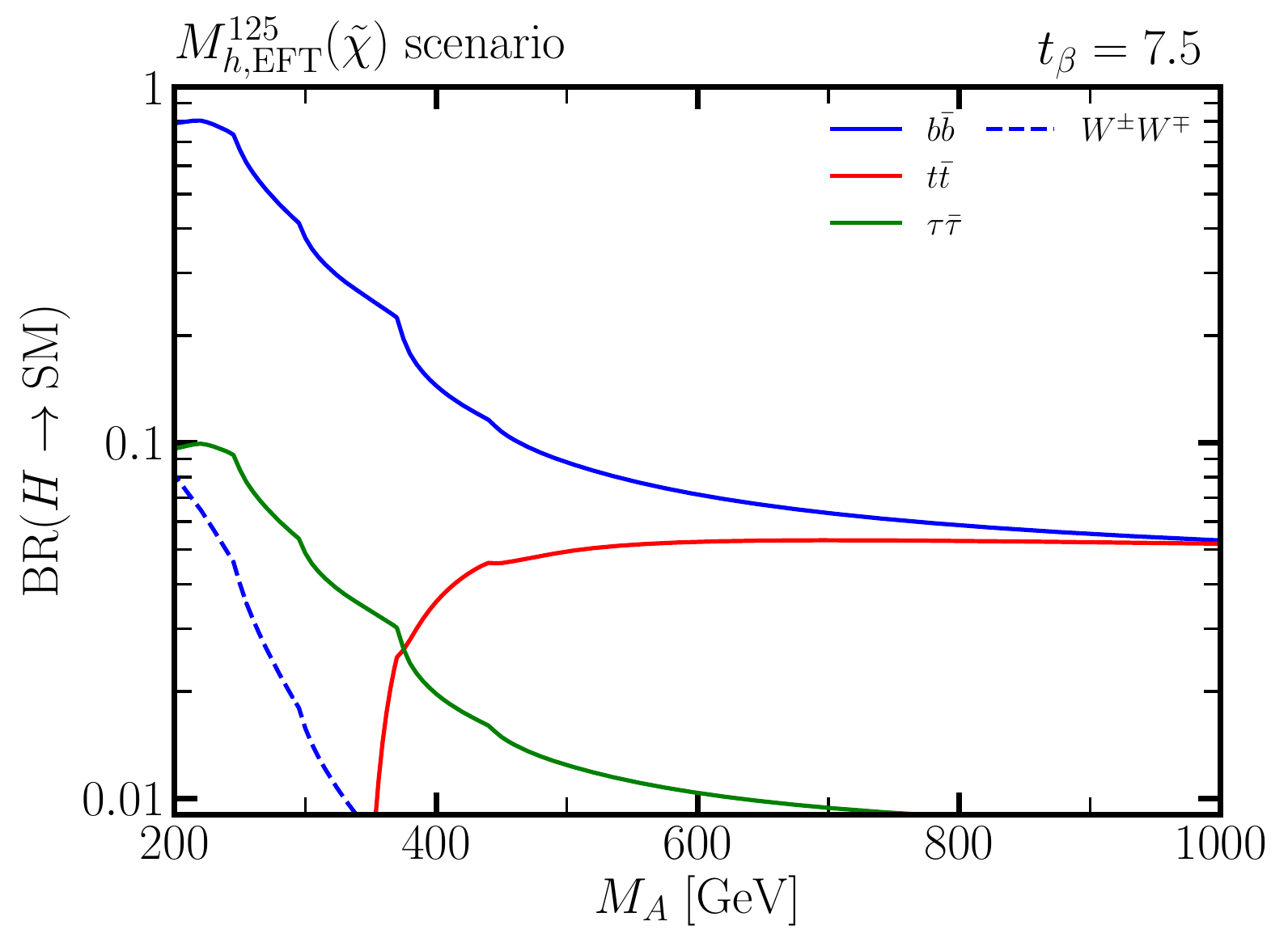}
\end{minipage}
\begin{minipage}{.5\textwidth}
\includegraphics[width=\textwidth]{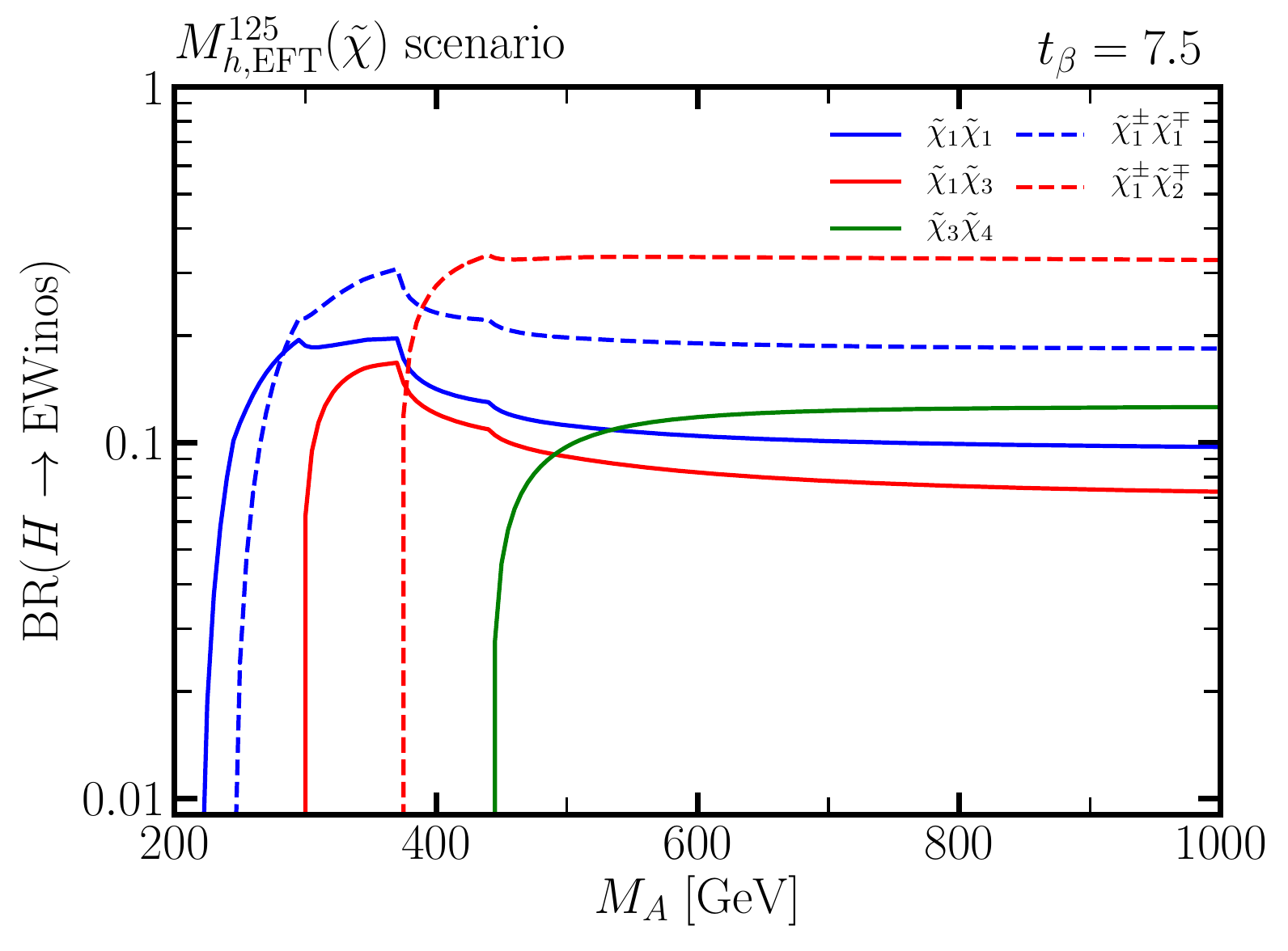}
\end{minipage}
\vspace*{-5mm}
\caption{Branching ratios for the decays of the heavy \cp-even Higgs boson $H$ in the \ltblchi scenario as a function of $M_A$, for fixed $\tb = 2.5$ (\emph{upper~panels}) and $7.5$ (\emph{lower~panels}). The dominant decays into \sm{} particles (\emph{left panels}) and electroweakinos (\emph{right panels}) are displayed.}
\label{fig:mh125-ltb-chi_BRH}
\end{figure}

To disentangle in detail the different decay channels of the heavy Higgs bosons, \fig{fig:mh125-ltb-chi_BRH} shows the branching ratios for the dominant decay modes of the \cp-even heavy Higgs boson $H$ into \sm{} particles (including the light Higgs boson $h$) (\emph{left panels}) and \susy{} particles (\emph{right panels}) as a function of $M_A$, focusing on the region $200\,\GeV \le M_A \le 1000$\,GeV, for $\tb=2.5$ (\emph{upper~panels}) and $\tb=7.5$ (\emph{lower~panels}). For $\tb = 2.5$ the decays into \sm{} particles dominate for $M_A\lesssim 350$\,GeV with the strongest decays modes being $H\to W^\pm W^\mp, ZZ$ (for $M_A\lesssim 250$\,GeV) and $H\to hh$ (for $M_A\gtrsim 250$\,GeV). In the $M_A$ range between $280$\,GeV and $360$\,GeV, the dominant $H$ decay to electroweakinos is $H\to \tilde{\chi}_1\tilde{\chi}_3$, featuring a decay rate of up to $30\%$. For $M_A\gtrsim 350$\,GeV, the decay into a pair of top quarks becomes kinematically accessible, reaching around $30\%$ at high $M_A$ values, and leading to a suppression of the other decay modes. In this $M_A$ range, all other relevant decays contain electroweakinos in the final state, with the dominant decay modes being $H\to\tilde\chi_1^\pm\tilde\chi_2^\mp$ ($\sim 30\%$), $H\to \tilde\chi_1\tilde\chi_3$ ($\sim10\%$) and $H\to \tilde\chi_3\tilde\chi_4$ ($\sim10\%$).

For $\tb = 7.5$, the $H$ decay into a top-quark pair is suppressed, while the decay into a bottom-quark pair and a $\tau$-lepton pair is enhanced. In particular, the latter decays play a significant role for $M_A \lesssim 380$\,GeV, whereas the $H$ decays to vector bosons or light Higgs bosons are negligible. In this low $M_A$ range, the dominant Higgs-to-electroweakino decays are to the lighter states, $H\to \tilde\chi_1^\pm\tilde\chi_1^\mp$, the invisible decay $H\to \tilde\chi_1\tilde\chi_1$ and $H\to \tilde\chi_1\tilde\chi_3$, reaching decay rates of around $30\%$, $20\%$ and $17\%$, respectively, at $M_A \simeq 380$\,GeV. At larger $M_A$ values, the $H$ decays are strongly dominated by the electroweakino final states, with $H\to\tilde\chi_1^\pm\tilde\chi_2^\mp$  ($\sim 30\%$),  $H\to\tilde\chi_1^\pm\tilde\chi_1^\mp$  ($\sim 20\%$),  $H\to \tilde\chi_1\tilde\chi_4$ ($\sim12\%$) and $H\to \tilde\chi_1\tilde\chi_1$ ($\sim10\%$) being the dominant decay modes. In general, for both $\tb$ values, we observe that the branching ratios are almost constant for $M_A\gtrsim 500$\,GeV, i.e.~as soon as all decays of $H$ into pairs of electroweakinos are kinematically open.

\begin{figure}[t]
\begin{minipage}{.5\textwidth}
\includegraphics[width=\textwidth]{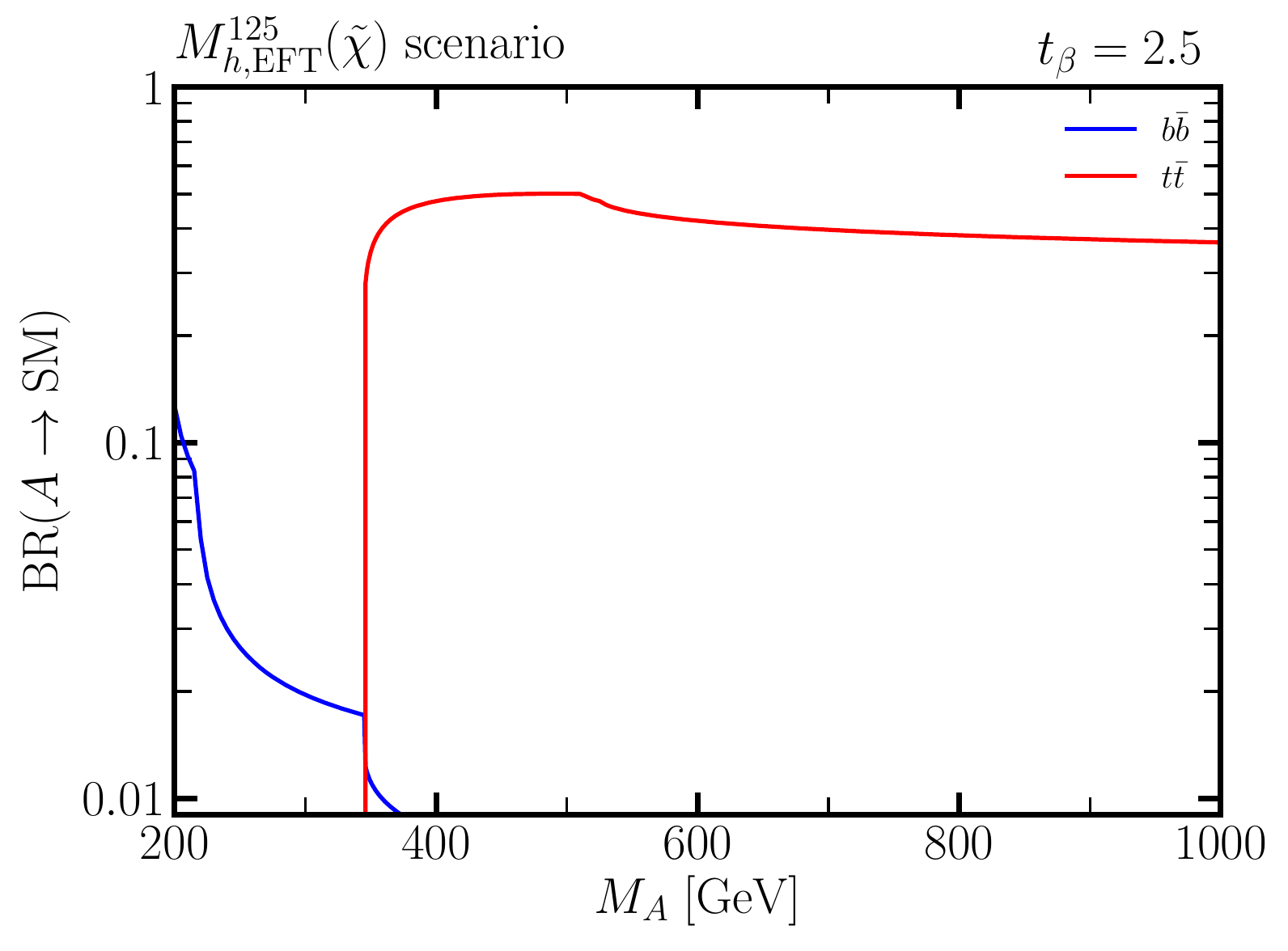}
\end{minipage}
\begin{minipage}{.5\textwidth}
\includegraphics[width=\textwidth]{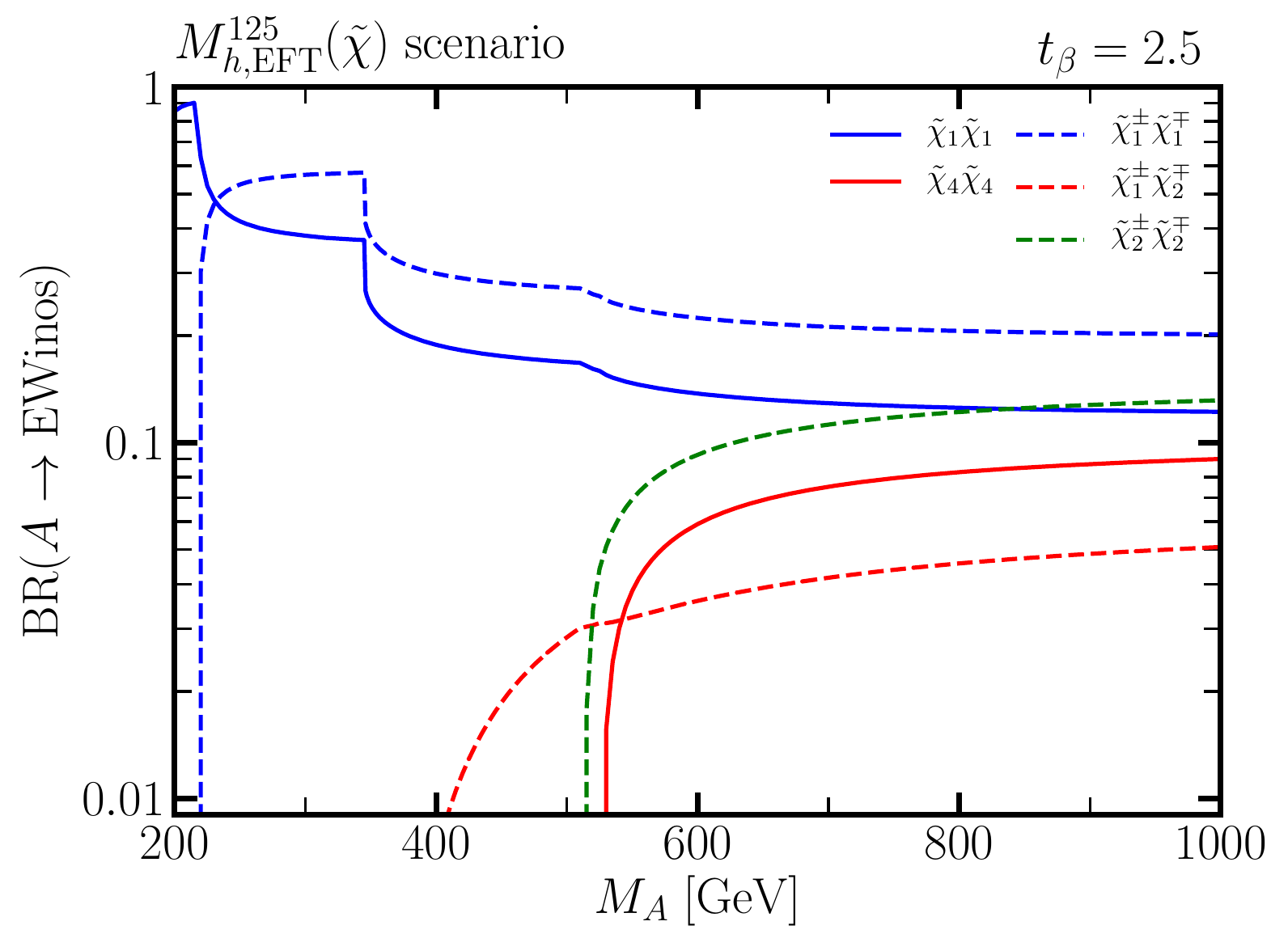}
\end{minipage}
\begin{minipage}{.5\textwidth}
\includegraphics[width=\textwidth]{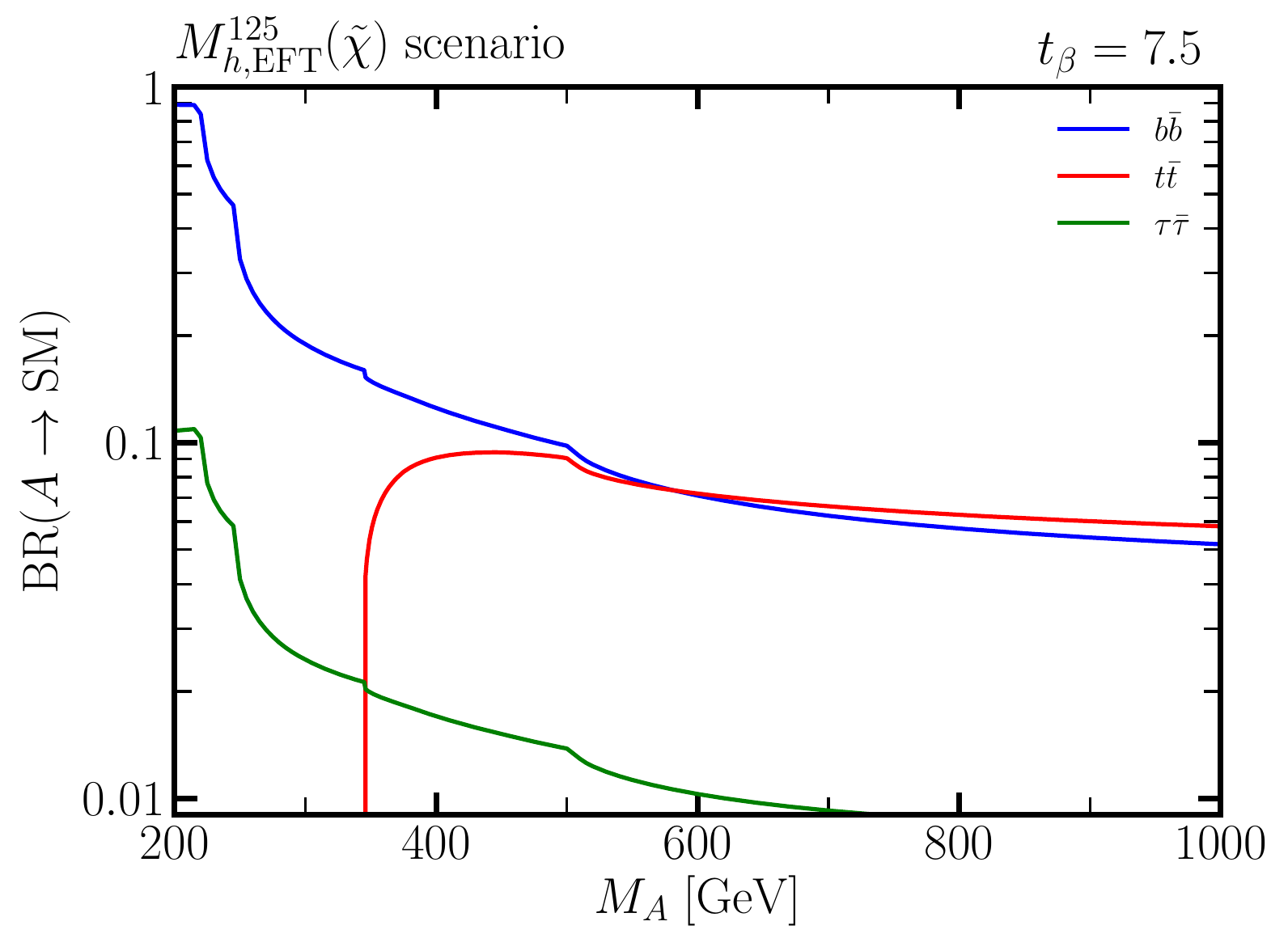}
\end{minipage}
\begin{minipage}{.5\textwidth}
\includegraphics[width=\textwidth]{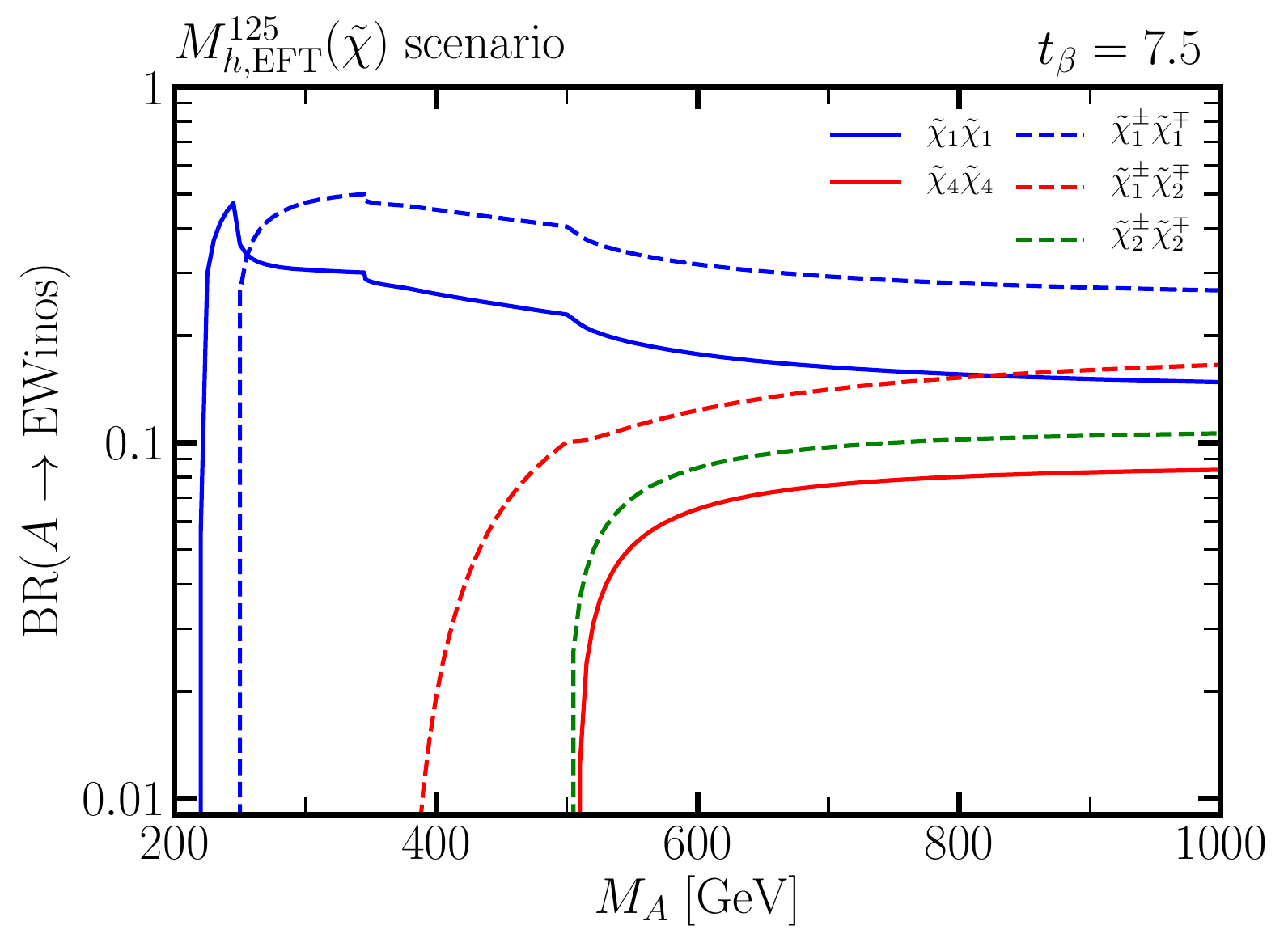}
\end{minipage}
\vspace*{-5mm}
\caption{Branching ratios for the decays of the \cp-odd Higgs boson $A$ in the \ltblchi scenario as a function of $M_A$, for fixed $\tb = 2.5$ (\emph{upper~panels}) and $7.5$ (\emph{lower~panels}). The dominant decays into \sm{} particles (\emph{left panels}) and electroweakinos (\emph{right panels}) are displayed.}
\label{fig:mh125-ltb-chi_BRA}
\end{figure}

We now turn to the decays of the \cp-odd Higgs boson $A$ in the \ltblchi scenario, which are shown in \fig{fig:mh125-ltb-chi_BRA} in analogy to \fig{fig:mh125-ltb-chi_BRH}. For $\tb=2.5$, the decays into \sm{} particles only play a minor role below the threshold for decays into a top-quark pair. In this region, the Higgs-to-electroweakino decays $A\to\tilde\chi_1\tilde\chi_1$ and $A\to\tilde\chi_1^\pm\tilde\chi_1^\mp$ are dominant. Above the top-quark pair threshold, the $A$~boson dominantly decays into $t\bar{t}$ with branching ratios of $\sim 40\%$ to $50\%$ and the decay modes into electroweakino pairs $A\to\tilde\chi_1^\pm\tilde\chi_1^\mp$ and $A\to\tilde\chi_1\tilde\chi_1$ drop to $\sim 20\%$ and $\sim 12\%$, respectively. Other electroweakino final states become accessible at higher $M_A$ values, however, they remain mostly subdominant.

For $\tb=7.5$, the pattern of $A$ decays to \sm{} particles changes significantly. Below the $A\to t\bar{t}$ kinematic threshold, the decay $A\to b\bar{b}$ dominates. Once the electroweakino final states become kinematically accessible for $M_A \gtrsim 220$\,GeV, the invisible decay $A\to\tilde\chi_1\tilde\chi_1$ and the decay $A\to\tilde\chi_1^\pm\tilde\chi_1^\mp$ reach values of up to $40$--$50\%$, thus leading to a strong suppression of $A$ decays to \sm{} final states. Once the decay $A\to\tilde\chi_1^\pm\tilde\chi_1^\mp$ is open, it remains the dominant decay mode with a rate of $\sim 30\%$ at high $M_A$ values. Here, the next-to-highest decays rates are obtained for $A\to \tilde\chi_1^\pm\tilde\chi_2^\mp$ and $A\to \tilde\chi_1 \tilde\chi_1$, amounting to around $10$--$20\%$ each. The decays to \sm{}~particles, i.e. $A\to t\bar{t}$ and $A\to b\bar{b}$, have a combined
branching ratio of $\sim11\%$ at large $M_A$.

\begin{figure}[t]
\begin{minipage}{.5\textwidth}
\includegraphics[width=\textwidth]{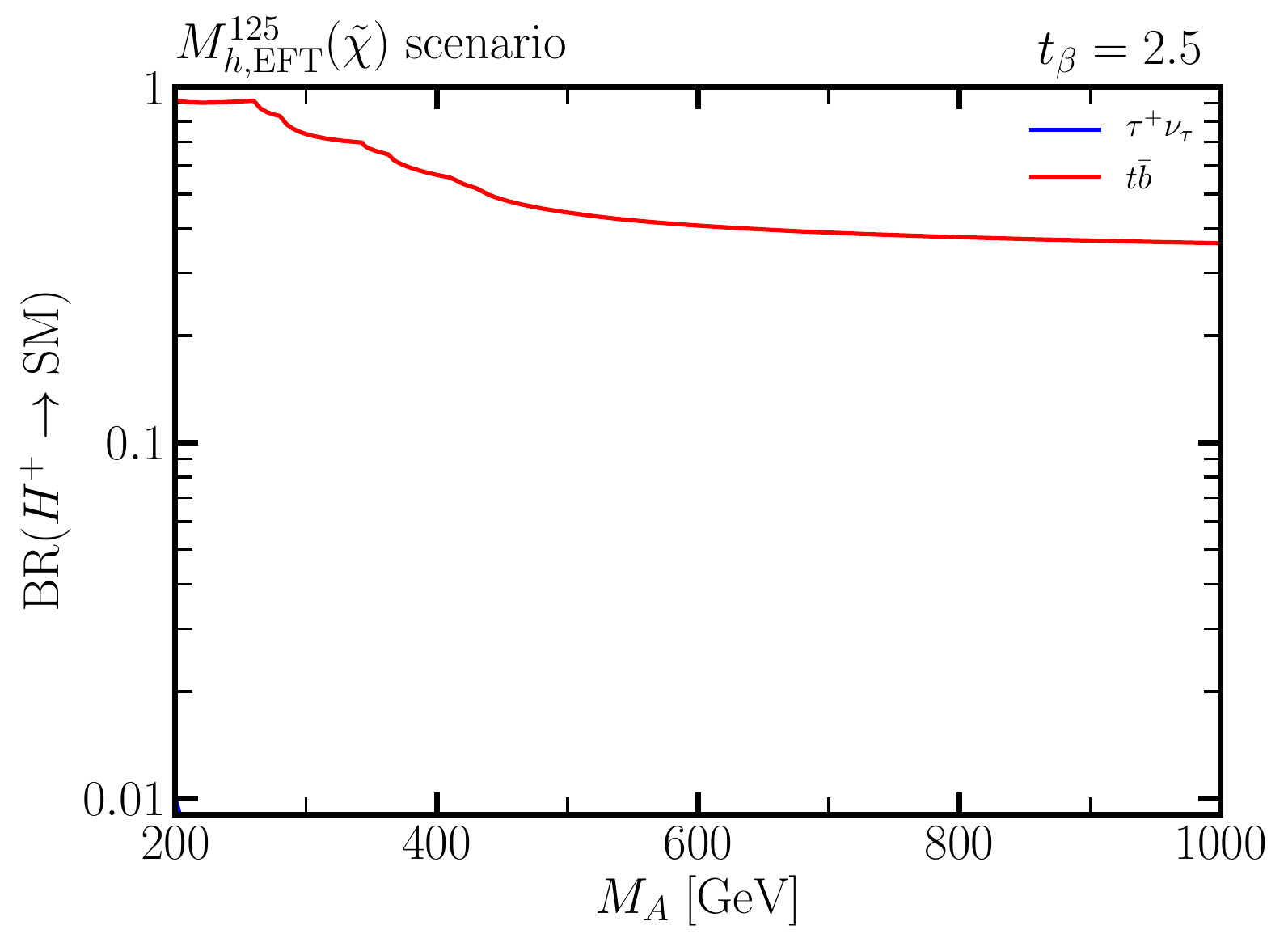}
\end{minipage}
\begin{minipage}{.5\textwidth}
\includegraphics[width=\textwidth]{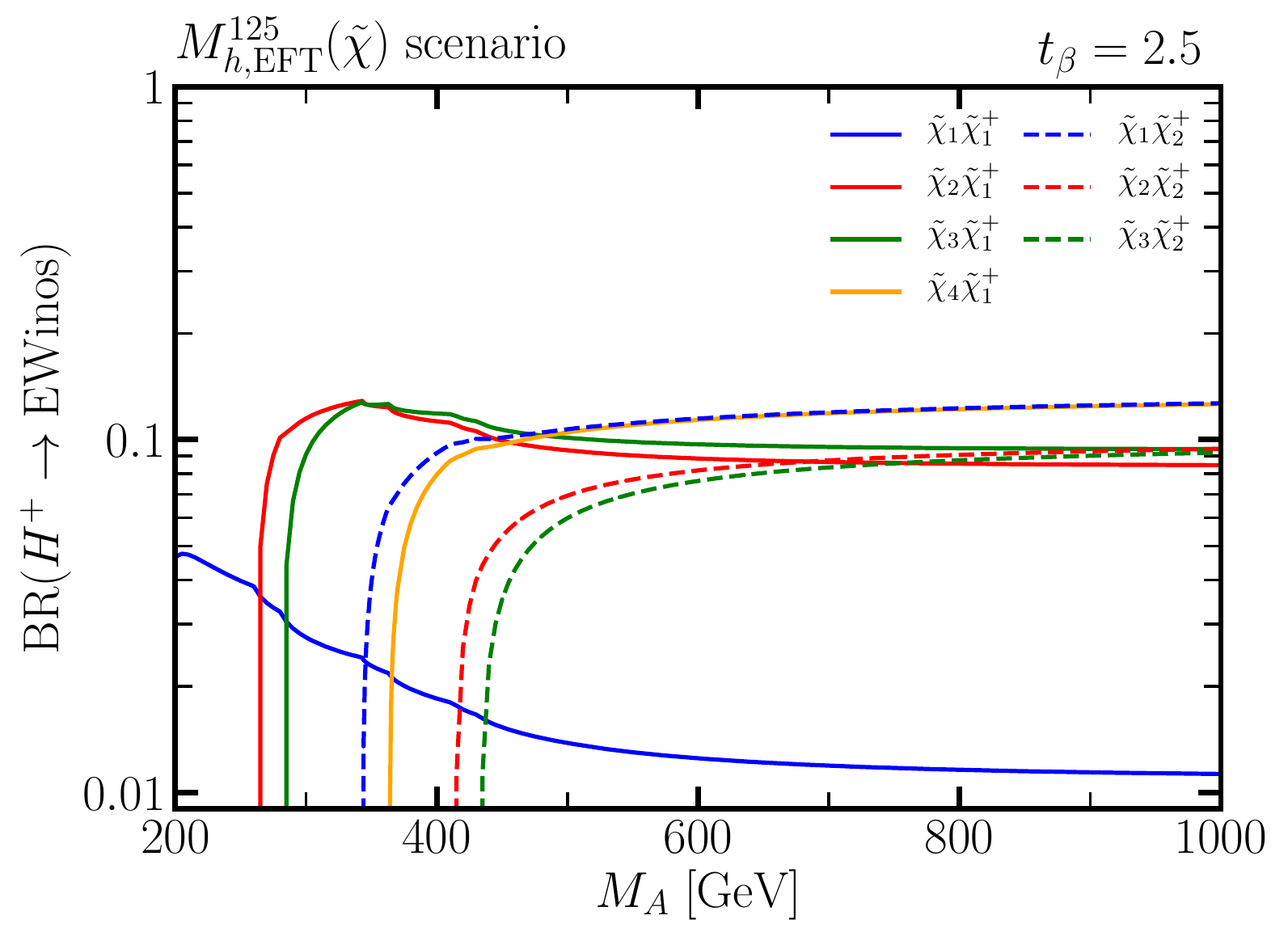}
\end{minipage}
\begin{minipage}{.5\textwidth}
\includegraphics[width=\textwidth]{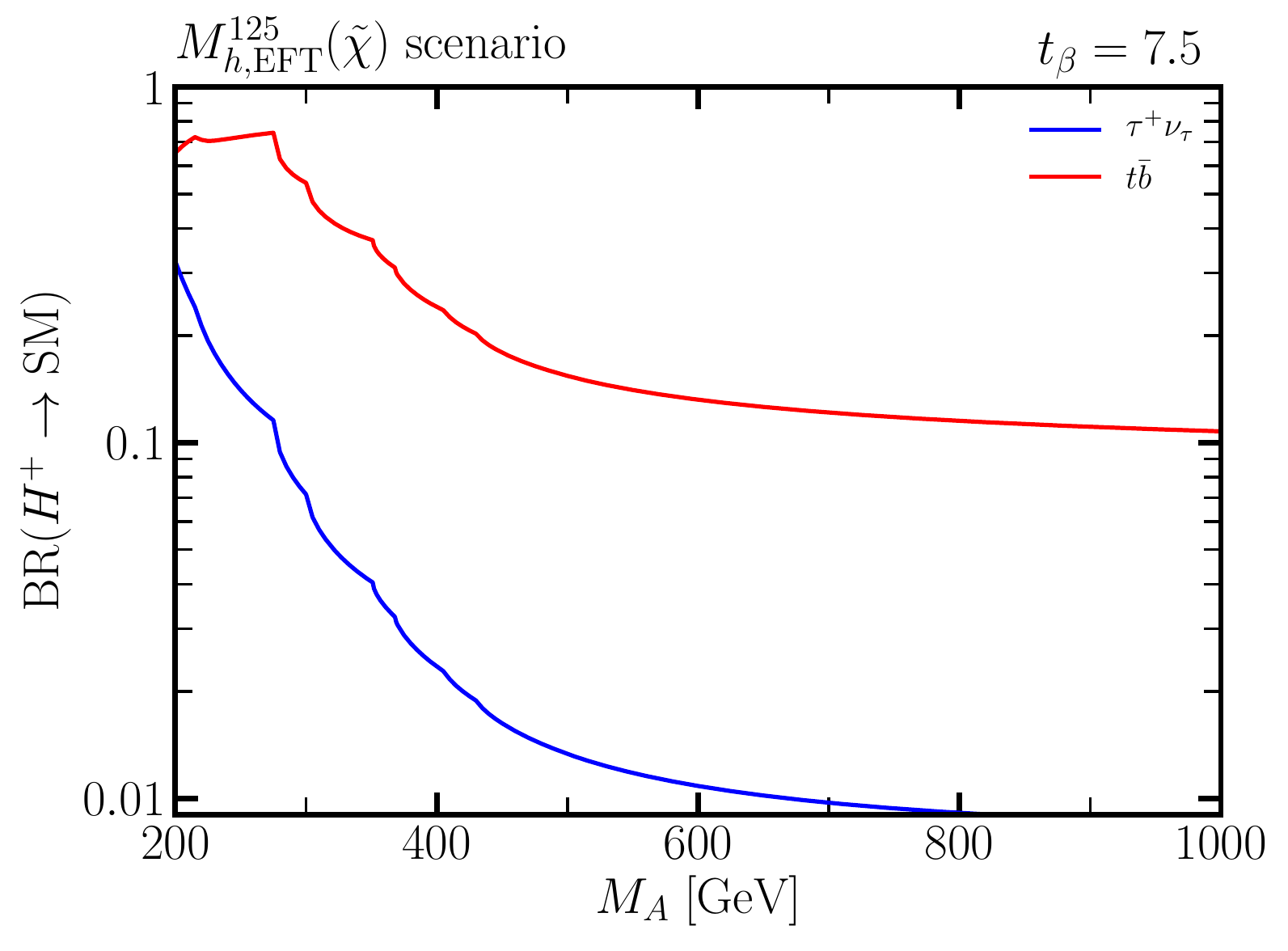}
\end{minipage}
\begin{minipage}{.5\textwidth}
\includegraphics[width=\textwidth]{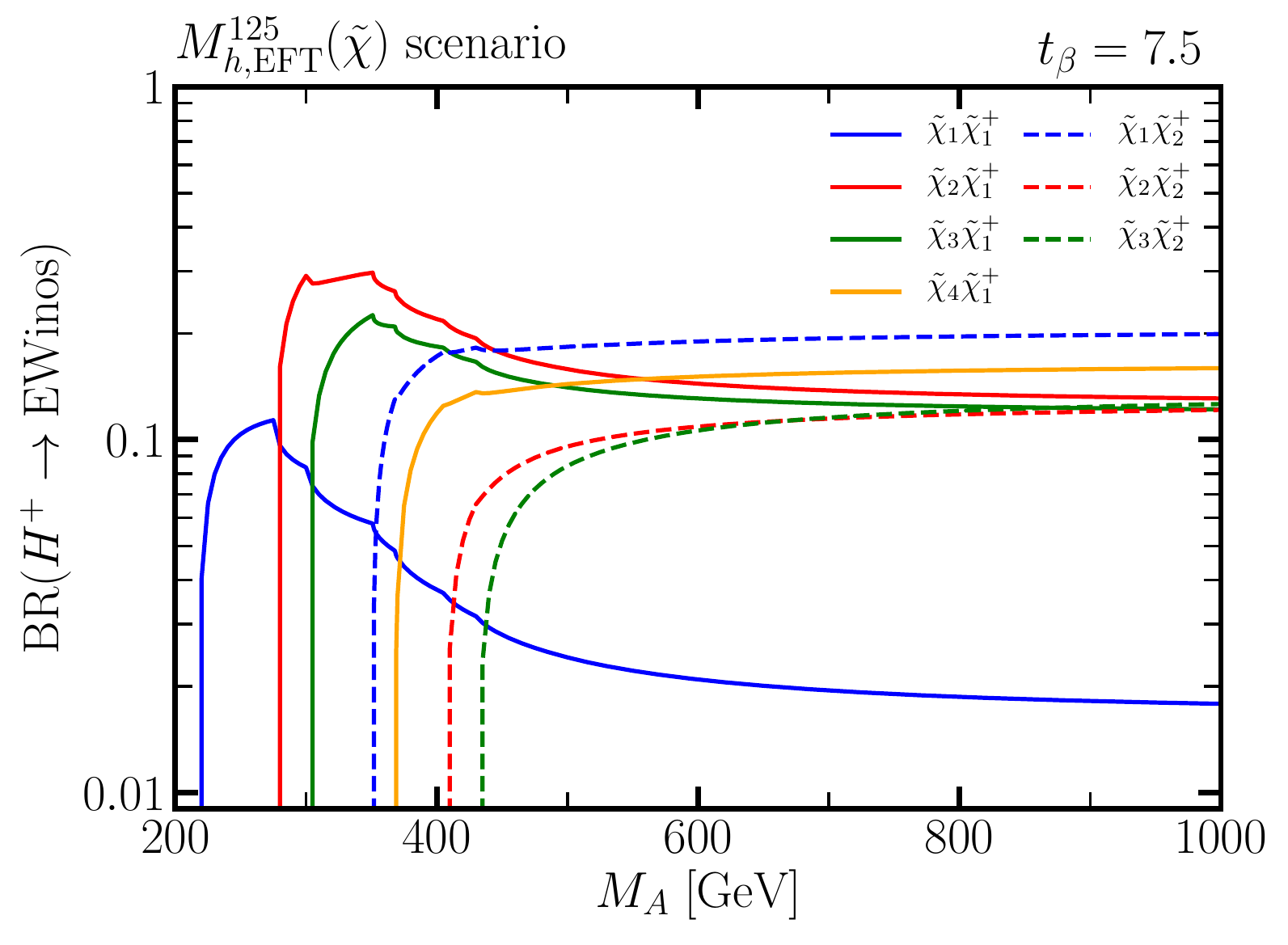}
\end{minipage}
\vspace*{-5mm}
\caption{Branching ratios for the decays of the charged Higgs boson $H^+$ in the \ltblchi scenario as a function of $M_A$, for fixed $\tb = 2.5$ (\emph{upper~panels}) and $7.5$ (\emph{lower~panels}). The dominant decays into \sm{} particles (\emph{left panels}) and electroweakinos (\emph{right panels}) are displayed.}
\label{fig:mh125-ltb-chi_BRHp}
\end{figure}

Lastly, we study in detail the decays of the charged Higgs boson in \fig{fig:mh125-ltb-chi_BRHp} in analogy to \figs{fig:mh125-ltb-chi_BRH} and \ref{fig:mh125-ltb-chi_BRA}. For $\tb = 2.5$, the dominant decay mode is $H^+\to t\bar b$, which does not fall below $35\%$ in the considered $M_A$ range. The decays into neutralino-chargino pairs, once they become kinematically accessible, are all comparable in size, with the strongest modes reaching branching ratios of $\sim 12\%$. For $\tb = 7.5$, the overall behavior of the various decay modes is very similar to the case of $\tb=2.5$. For $M_A\lesssim 300$\,GeV, however, the branching ratio into $t\bar b$ is reduced by $\sim 20$--$30\%$, while the decay to $\tau^+\nu_\tau$ is much more important, with decay rates of up to $30\%$. Also the branching ratios for decays into neutralino-chargino pairs is increased, with $\mathrm{BR}(H^+\to \tilde\chi_2\tilde\chi_1^\pm)$ reaching values up to $30\%$ for $M_A \sim (280$--$360)$\,GeV. The fact that the rates of various $H^+$-to-electroweakino decays are very similar warrants a combined experimental search for the various signatures arising from these decays within this benchmark model.

The previous figures clearly show that decay rates of heavy neutral and charged Higgs bosons to electroweakinos can be sizable in the \ltblchi scenario. This strongly motivates dedicated searches for these type of decays. In the remaining part of this section we will therefore discuss such decays in more detail, focusing on a few representative parameter points.

Typically, \lhc{} searches for electroweakino pairs that are produced via the conventional processes, namely $s$-channel vector-boson exchange and $t$-channel squark exchange, and decay to the lightest supersymmetric particle (\lsp{}) --- the lightest neutralino, $\tilde\chi_1$ --- by emitting a $W$- or a $Z$-boson, select events with $\ge 2$ leptons (i.e., electrons and muons) and large missing transverse energy, $\etmiss$, see e.g.~Refs.~\cite{Sirunyan:2017lae,Sirunyan:2018lul,Sirunyan:2018ubx,Aaboud:2018sua,Aaboud:2018zeb,Aaboud:2018jiw} for recent \lhc{} searches. However, their sensitivity is significantly deteriorated when the mass difference between the decaying electroweakino and the \lsp{} is small. In this case, the emitted off-shell vector boson can only yield a low-$p_T$ (``soft'') lepton that often does not pass the lepton-reconstruction criteria. Dedicated searches for such compressed electroweakino mass spectra have been designed~\cite{Sirunyan:2018iwl,Aaboud:2017leg}, which require additional jet(s) from initial-state radiation (ISR) against which the produced electroweakinos recoil, thus giving an additional boost to the final-state leptons. However, these searches pay the price of a lower expected signal-event yield due to the ISR jet(s) requirement, and can therefore only cover the parameter space with very light electroweakinos.

The \ltblchi scenario features a heavily-mixed electroweakino sector, such that the limits obtained by the direct \lhc{} searches mentioned above cannot be directly applied. These are obtained under certain simplifying assumptions, e.g., for the stronger, ``pure wino'' production scenario of Refs.~\cite{Sirunyan:2018iwl,Aaboud:2017leg}: (i) the produced neutralinos and charginos are pure winos, (ii) their masses are equal, and (iii) they decay to $100\%$ into the lightest neutralino and a $Z$- or a $W$-boson, respectively. If these assumptions are fulfilled, current results with $36\,\mathrm{fb}^{-1}$ of $13$\,TeV data exclude ``wino'' masses of around $180$--$230$\,GeV for mass differences to the \lsp{} between $5$\,GeV and $25$\,GeV. We expect these searches to also exhibit some sensitivity to the electroweakino spectrum of the \ltblchi scenario, however, it is unclear whether they indeed exclude this scenario, given the complexity of its mass and decay spectrum (see below for details). We therefore strongly encourage dedicated analyses optimized for our and similar scenarios by the experiments.

\begin{table}[ht!]
\begin{center}
\small
\begin{tabular}{| c c | c | c | c | c |}
\toprule
\multicolumn{2}{|c|}{\hfill} & scenario 1 & scenario 2 & scenario 3 & scenario 4 \\ 
\hline\hline
\multicolumn{2}{|c|}{$\mu\;[\GeV]$} & 180 & 180 & 280 & 280 \\
\rowcolor{Gray}\multicolumn{2}{|c|}{$M_1\;[\GeV]$} & 160 & 160 & 260 & 260 \\
\multicolumn{2}{|c|}{$M_2\;[\GeV]$} & 180 & 180 & 280 & 280 \\
\rowcolor{Gray}\multicolumn{2}{|c|}{$\tb$}   & 2.5 & 7.5 & 2.5 & 7.5 \\
\hline\hline
\multicolumn{6}{|c|}{Neutralino and chargino masses, production and decay rates} \\
\hline\hline
\multicolumn{2}{|c|}{$m_{\tilde\chi_1}\;[\GeV]$}     & 95.8  & 110.0 & 194.8 & 207.9 \\
\rowcolor{Gray}\multicolumn{2}{|c|}{$m_{\tilde\chi_2}\;[\GeV]$}     & 164.6 & 164.7 & 264.6 & 254.7 \\
\multicolumn{2}{|c|}{$m_{\tilde\chi_3}\;[\GeV]$}     & 183.8 & 188.8 & 282.4 & 285.6 \\
\rowcolor{Gray}\multicolumn{2}{|c|}{$m_{\tilde\chi_4}\;[\GeV]$}     & 263.4 & 254.1 & 362.9 & 353.0 \\
\multicolumn{2}{|c|}{$m_{\tilde\chi^\pm_1}\;[\GeV]$} & 108.9 & 122.7 & 207.9 & 220.4 \\
\rowcolor{Gray}\multicolumn{2}{|c|}{$m_{\tilde\chi^\pm_2}\;[\GeV]$} & 256.6 & 250.3 & 355.7 & 348.1 \\
\hline
\multicolumn{1}{|l}{\hspace{0.1cm} $\sigma(pp \to \tilde\chi_1 \tilde\chi_3)$ } & \multicolumn{1}{l|}{[fb]} &	407.0	&	329.0	&	58.7		&		52.3		\\
\rowcolor{Gray}\multicolumn{1}{|l}{\hspace{0.1cm} $\sigma(pp \to \tilde\chi_3 \tilde\chi_4)$ } & \multicolumn{1}{l|}{[fb]} &	77.1	&	77.1	&	18.8		&		19.2		\\
\multicolumn{1}{|l}{\hspace{0.1cm} $\sigma(pp \to \tilde\chi_1^\pm \tilde\chi_1)$ } & \multicolumn{1}{l|}{[fb]} &	8661.0	&	5206.0	&	754.0		&		579.4		\\
\rowcolor{Gray}\multicolumn{1}{|l}{\hspace{0.1cm} $\sigma(pp \to \tilde\chi_1^\pm \tilde\chi_2)$ } & \multicolumn{1}{l|}{[fb]} &	974.0	&	866.0	&	135.0		&		128.3		\\
\multicolumn{1}{|l}{\hspace{0.1cm} $\sigma(pp \to \tilde\chi_1^\pm \tilde\chi_3)$ } & \multicolumn{1}{l|}{[fb]} &	660.0	&	547.0	&	102.0		&		91.9		\\
\rowcolor{Gray}\multicolumn{1}{|l}{\hspace{0.1cm} $\sigma(pp \to \tilde\chi_1^\pm \tilde\chi_4)$ } & \multicolumn{1}{l|}{[fb]} &	87.7	&	99.0	&	18.7		&		20.0		\\

\multicolumn{1}{|l}{\hspace{0.1cm} $\sigma(pp \to \tilde\chi_2^\pm \tilde\chi_2)$ } & \multicolumn{1}{l|}{[fb]} &	132.0	&	136.0	&	31.0		&		31.5		\\
\rowcolor{Gray}\multicolumn{1}{|l}{\hspace{0.1cm} $\sigma(pp \to \tilde\chi_2^\pm \tilde\chi_3)$ } & \multicolumn{1}{l|}{[fb]} &	154.0	&	160.0	&	37.3		&		38.7		\\
\multicolumn{1}{|l}{\hspace{0.1cm} $\sigma(pp \to \tilde\chi_2^\pm \tilde\chi_4)$ } & \multicolumn{1}{l|}{[fb]} &	331.2	&	371.0	&	92.9		&		102.3		\\
\rowcolor{Gray}\multicolumn{1}{|l}{\hspace{0.1cm} $\sigma(pp \to \tilde\chi_1^\pm \tilde\chi_1^\mp)$ } & \multicolumn{1}{l|}{[fb]} &	4613.0	&	2999.7	&	440.0		&		352.2		\\
\multicolumn{1}{|l}{\hspace{0.1cm} $\sigma(pp \to \tilde\chi_1^\pm \tilde\chi_2^\mp)$ } & \multicolumn{1}{l|}{[fb]} &	71.4	&	65.6	&	15.2		&		14.3		\\
\rowcolor{Gray}\multicolumn{1}{|l}{\hspace{0.1cm} $\sigma(pp \to \tilde\chi_2^\pm \tilde\chi_2^\mp)$ } & \multicolumn{1}{l|}{[fb]} &	199.0	&	217.2	&	54.5		&		58.9		\\
\hline
\multicolumn{2}{|c|}{$\mathrm{BR}(\tilde\chi_2 \to \dots)$} &
100.0\%  ($\tilde\chi_1^\pm W^{\mp\,*}$) &
99.0\% ($\tilde\chi_1^\pm W^{\mp\,*}$)&
100\% ($\tilde\chi_1^\pm W^{\mp\,*}$)&
100\% ($\tilde\chi_1^\pm W^{\mp\,*}$)\\
\rowcolor{Gray}
\multicolumn{2}{|c|}{} &
54.8\% ($\tilde\chi_1^\pm W^{\mp\,*}$) &
55.5\% ($\tilde\chi_1^\pm W^{\mp\,*}$) &
52.2\% ($\tilde\chi_1^\pm W^{\mp\,*}$) &
52.4\% ($\tilde\chi_1^\pm W^{\mp\,*}$) \\
\rowcolor{Gray}
\multicolumn{2}{|c|}{\multirow{-2}{*}{$\mathrm{BR}(\tilde\chi_3 \to \dots)$}} &
45.2\% ($\tilde\chi_1^0 Z^{*}$) &
44.5\% ($\tilde\chi_1^0 Z^{*}$) &
47.8\% ($\tilde\chi_1^0 Z^{*}$) &
47.6\% ($\tilde\chi_1^0 Z^{*}$) \\
\multicolumn{2}{|c|}{$\mathrm{BR}(\tilde\chi_4^0 \to \dots)$} &
99.5\% ($\tilde\chi_1^\pm W^{\mp}$) &
99.1\% ($\tilde\chi_1^\pm W^{\mp}$) &
99.8\% ($\tilde\chi_1^\pm W^{\mp}$) &
99.5\% ($\tilde\chi_1^\pm W^{\mp}$) \\
\rowcolor{Gray}
\multicolumn{2}{|c|}{$\mathrm{BR}(\tilde\chi_1^\pm \to \dots)$} &
100\% ($\tilde\chi_1^0 W^{\pm\,*}$) &
100\% ($\tilde\chi_1^0 W^{\pm\,*}$) &
100\% ($\tilde\chi_1^0 W^{\pm\,*}$) &
100\% ($\tilde\chi_1^0 W^{\pm\,*}$)\\
\multicolumn{2}{|c|}{\multirow{2}{*}{$\mathrm{BR}(\tilde\chi_2^\pm \to \dots)$}} &
53.4\% ($\tilde\chi_1^\pm Z$) &
51.5\% ($\tilde\chi_1^\pm Z$) &
53.7\% ($\tilde\chi_1^\pm Z$) &
52.7\% ($\tilde\chi_1^\pm Z$) \\
\multicolumn{2}{|c|}{} &
38.0\% ($\tilde\chi_1^0 W^{\pm}$) &
41.8\% ($\tilde\chi_1^0 W^{\pm}$) &
40.1\% ($\tilde\chi_1^0 W^{\pm}$) &
44.7\% ($\tilde\chi_1^0 W^{\pm}$)\\
\bottomrule
\end{tabular}
\end{center}
\vspace*{-6mm}
\caption{A detailed view on two parameter points from the \ltblchi scenario (scenario 1 and 2), as well as two variations (scenario 3 and 4): relevant parameters for the electroweak sector (\emph{top panel}); masses and rates for the dominant production modes (for the \lhc{} at $13$\,TeV)  and decay modes of the neutralinos and charginos (\emph{bottom panel}).}
\label{tab:EWino_scenarios1}
\end{table}

\begin{table}[ht]
\begin{center}
\small
\begin{tabular}{| c c | c | c | c | c |}
\toprule
\multicolumn{2}{|c|}{\hfill} & scenario 1 & scenario 2 & scenario 3 & scenario 4 \\ 
\hline\hline
\multicolumn{2}{|c|}{$\mu\;[\GeV]$} & 180 & 180 & 280 & 280 \\
\rowcolor{Gray}\multicolumn{2}{|c|}{$M_1\;[\GeV]$} & 160 & 160 & 260 & 260 \\
\multicolumn{2}{|c|}{$M_2\;[\GeV]$} & 180 & 180 & 280 & 280 \\
\rowcolor{Gray}\multicolumn{2}{|c|}{$\tb$}   & 2.5 & 7.5 & 2.5 & 7.5 \\
\hline\hline
\multicolumn{6}{|c|}{\mssm{} Higgs boson production and decay rates for $M_A = 1$\,TeV } \\
\hline\hline
\rowcolor{Gray}\multicolumn{1}{|l}{\hspace{0.7cm}$\sigma(gg\to H)$    } & \multicolumn{1}{l|}{[fb]} & 19.0 & 1.8 & 19.0 & 1.8 \\
\multicolumn{1}{|l}{\hspace{0.7cm}$\sigma(b\bar{b}\to H)$   } & \multicolumn{1}{l|}{[fb]} &  0.6 & 5.1 &  0.6 &  5.1 \\
\rowcolor{Gray}\multicolumn{1}{|l}{\hspace{0.7cm}$\sigma(gg\to A)$    } & \multicolumn{1}{l|}{[fb]} & 24.6 & 3.6 & 24.6 & 3.6 \\
\multicolumn{1}{|l}{\hspace{0.7cm}$\sigma(b\bar{b}\to A)$   } & \multicolumn{1}{l|}{[fb]} &  0.6 & 5.1 &  0.6 &  5.1 \\
\rowcolor{Gray}\multicolumn{1}{|l}{\hspace{0.7cm}$\sigma(pp\to t H^-)$} & \multicolumn{1}{l|}{[fb]} &  3.7 & 0.7 &  3.7 &  0.7 \\
\hline
\hfill & \hfill & 33.1\% ($\tilde\chi_1^\pm\tilde\chi_2^\mp$) & 32.6\% ($\tilde\chi_1^\pm\tilde\chi_2^\mp$) & 33.0\% ($\tilde\chi_1^\pm\tilde\chi_2^\mp$) & 36.0\% ($\tilde\chi_1^\pm\tilde\chi_2^\mp$) \\
\multicolumn{2}{|c|}{$\mathrm{BR}(H\to\tilde\chi\tilde\chi)$} & 11.1\% ($\tilde\chi_3\tilde\chi_4$) & 18.4\% ($\tilde\chi_1^\pm\tilde\chi_1^\mp$) & 10.9\% ($\tilde\chi_1\tilde\chi_3$) & 14.4\% ($\tilde\chi_1^\pm\tilde\chi_1^\mp$) \\
\hfill & \hfill &
9.8\% ($\tilde\chi_1\tilde\chi_3$) & 12.6\% ($\tilde\chi_3\tilde\chi_4$) & 10.4\% ($\tilde\chi_3\tilde\chi_4$) & 12.4\% ($\tilde\chi_3\tilde\chi_4$) \\
\rowcolor{Gray}\hfill & \hfill &
20.2\% ($\tilde\chi_1^\pm\tilde\chi_1^\mp$) & 26.8\% ($\tilde\chi_1^\pm\tilde\chi_1^\mp$) & 19.6\% ($\tilde\chi_1^\pm\tilde\chi_1^\mp$) & 26.1\% ($\tilde\chi_1^\pm\tilde\chi_1^\mp$) \\
\rowcolor{Gray} \multicolumn{2}{|c|}{$\mathrm{BR}(A\to\tilde\chi\tilde\chi)$} & 13.2\% ($\tilde\chi_2^\pm\tilde\chi_2^\mp$) & 16.6\% ($\tilde\chi_1^\pm\tilde\chi_2^\mp$) & 12.5\% ($\tilde\chi_2^\pm\tilde\chi_2^\mp$) & 14.0\% ($\tilde\chi_1^\pm\tilde\chi_2^\mp$) \\
\rowcolor{Gray} \hfill & \hfill & 12.2\% ($\tilde\chi_1\tilde\chi_1$) & 14.8\% ($\tilde\chi_1\tilde\chi_1$) & 12.2\% ($\tilde\chi_1\tilde\chi_1$) & 15.2\% ($\tilde\chi_1\tilde\chi_1$) \\
\hfill & \hfill & 12.7\% ($\tilde\chi_1\tilde\chi_2^+$) & 19.9\% ($\tilde\chi_1\tilde\chi_2^+$) & 12.3\% ($\tilde\chi_4\tilde\chi_1^+$) & 17.4\% ($\tilde\chi_1\tilde\chi_2^+$) \\
\multicolumn{2}{|c|}{$\mathrm{BR}(H^+\to\tilde\chi\tilde\chi^+)$} & 12.6\% ($\tilde\chi_4\tilde\chi_1^+$) & 16.0\% ($\tilde\chi_4\tilde\chi_1^+$) & 11.2\% ($\tilde\chi_1\tilde\chi_2^+$) & 16.6\% ($\tilde\chi_4\tilde\chi_1^+$) \\
\hfill & \hfill & 9.4\% ($\tilde\chi_2\tilde\chi_2^+$) & 13.1\% ($\tilde\chi_2\tilde\chi_1^+$) & 9.5\% ($\tilde\chi_3\tilde\chi_1^+$) & 13.1\% ($\tilde\chi_3\tilde\chi_1^+$) \\
\hline
\rowcolor{Gray}  \multicolumn{2}{|c|}{$\mathrm{BR}(h\to\gamma\gamma)_{\text{\mssm{}}/\text{\sm{}}}$} & 1.12 & 1.02 & 1.02 & 0.98 \\
\bottomrule
\end{tabular}
\end{center}
\vspace*{-6mm}
\caption{A detailed view on two parameter points from the \ltblchi scenario (scenario 1 and 2), as well as two variations (scenario 3 and 4): relevant parameters for the electroweak sector (\emph{top panel}); rates of the dominant production modes (for the \lhc{} at $13$\,TeV) and decay modes to electroweakinos of the heavy Higgs bosons $H$, $A$ and $H^+$, for fixed $M_A = 1$\,TeV (\emph{bottom panel}).}
\label{tab:EWino_scenarios2}
\end{table}

In \tab{tab:EWino_scenarios1} we provide detailed information on the masses and dominant production and decay modes of the neutralinos and charginos in the \ltblchi benchmark scenario, for $\tb$ values of $2.5$ ({scenario~1}) and $7.5$ ({scenario~2}). We calculated the direct neutralino/chargino production cross sections for the \lhc{} at a center-of-mass energy of $13$\,\TeV at the \nlo{}+\nll{} level using \texttt{Resummino} (version 2.0.1)~\cite{Debove:2009ia,Debove:2010kf,Debove:2011xj,Fuks:2012qx,Fuks:2013vua} with the CT14 \pdf{} sets~\cite{Dulat:2015mca}. The value of $M_A$ does not affect the electroweakino spectrum at tree-level, however, it obviously affects the heavy Higgs-boson phenomenology. In \tab{tab:EWino_scenarios2} we provide for these scenarios the $13$\,TeV cross sections for the dominant heavy Higgs-boson production modes, as well as the rates for the three dominant decays to electroweakinos, for $M_A = 1$\,TeV.

For the heavy \cp-even Higgs boson $H$ we identify the cascade decay
\begin{align}
H \rightarrow \tilde\chi_1^\pm \tilde\chi_2^\mp \rightarrow (\tilde \chi_1 W^{\pm\,*})(\tilde\chi_1^\mp Z) \rightarrow \tilde\chi_1\tilde\chi_1 W^{\pm\,*}W^{\mp\,*} Z
\end{align}
as the most frequent process, with a total rate of around $17.7\%$ ($16.8\%$) for $\tb = 2.5~(7.5)$. Despite their off-shellness, one can still expect reasonably high-$p_T$ leptons from the $W$-bosons, provided that $M_H \gg  (m_{\tilde\chi_1^\pm} + m_{\tilde\chi_2^\pm})$, as is the case in this example. This process can therefore lead to a spectacular signature with up to $4$ reconstructable leptons, missing transverse energy, and for larger $\tb$ values possibly two additional $b$-jets, if the heavy Higgs boson is produced in association with bottom quarks. Moreover, many of the other possible cascade decays also lead to final states with multiple $W$- and or $Z$-bosons. In contrast, the direct (invisible) Higgs-boson decay into two lightest neutralinos, $H\to \tilde\chi_1\tilde\chi_1$, as well as decays leading to a $Z + \etmiss$ final state (e.g., via $H\to \tilde{\chi}_1\tilde{\chi}_3 \to \tilde{\chi}_1\tilde{\chi}_1 Z$) occur with smaller rates, e.g.\ with branching ratios of $3.4\%$ and $4.4\%$, respectively, in scenario~1.\footnote{Our scenario(s) are therefore phenomenologically very different to those considered in Ref.~\cite{Gori:2018pmk}, where the final state $Z+\etmiss$ is regarded as the most promising search channel.}

For the \cp-odd Higgs boson $A$, the most frequent process is
\begin{align}
A \rightarrow \tilde\chi_1^\pm\tilde\chi_1^\mp \rightarrow \tilde\chi_1\tilde\chi_1 W^{\pm\,*} W^{\mp\,*}
\end{align}
with a rate of $20.2\%$ ($26.5\%$) for $\tb = 2.5~(7.5)$. Experimentally more promising, however, might be the cascade
\begin{align}
A \rightarrow \tilde \chi_2^\pm \tilde \chi_2^\mp \to (\tilde \chi_1^\pm Z)( \tilde \chi_1^\mp Z) \to \tilde\chi_1 \tilde\chi_1 Z Z W^{\pm\,*} W^{\mp\,*}
\end{align}
occurring with a branching ratio of $3.8\%$ ($2.8\%$), or
\begin{align}
A \rightarrow \tilde \chi_1^\pm \tilde \chi_2^\mp \to (\tilde \chi_1 W^{\pm\,*}) (\tilde \chi_1^\mp Z) \to \tilde\chi_1 \tilde\chi_1 Z W^{\pm\,*} W^{\mp\,*}
\end{align}
with a decay rate of $2.7\%$ ($8.5\%$)  for scenario 1 (scenario 2), i.e. $\tb=2.5~(7.5)$.

Finally, the most frequent charged Higgs-to-electroweakino decay cascade is
\begin{align}
H^\pm \to \tilde\chi_1 \tilde \chi_2^\pm \to \tilde\chi_1 (\tilde\chi_1^\pm Z) \to \tilde\chi_1 \tilde\chi_1 Z W^{\pm\,*}
\end{align}
with a rate of $6.8\%$ ($10.2\%$) in scenario 1 (scenario 2). Furthermore, many other possible charged Higgs-boson cascade decays yield final states with one or three $W$-bosons and missing transverse energy.

We stress again that searches for heavy Higgs-to-electroweakino processes are highly complementary to direct electroweakino searches, in particular in the case of a compressed electroweakino mass spectrum. While the final-state leptons in events from direct electroweakino production tend to be soft and difficult to reconstruct, this is not a problem in events where the electroweakinos originate from a heavy Higgs boson, and therefore come with a larger initial momentum. At the same time, the presence of a compressed electroweakino spectrum implies that multiple light electroweakino states are available, such that heavy Higgs-boson cascade decays via these states are possible (and even sometimes preferred, as demonstrated above), thus yielding multiple $W$- and $Z$-bosons in the final state.

For the current choice of electroweakino masses, and the example value of $M_A = 1\;\TeV$ in \tab{tab:EWino_scenarios2}, the production cross sections for direct electroweakino-pair production exceed the heavy Higgs-boson production cross sections by roughly two orders of magnitude. The leading direct electroweakino-production channels are the chargino-pair production processes $pp\to \tilde\chi^\pm_1\tilde\chi^\mp_1$ and $pp\to \tilde\chi^\pm_2\tilde\chi^\mp_2$. Due to the large difference in the rates, a dedicated search analysis of direct electroweakino production might turn out to be more sensitive than a dedicated search for the heavy Higgs-to-electroweakino channels for this specific parameter choice, despite the more experimentally challenging kinematics in the first case. However, lifting the electroweakino-mass spectrum (see below) and/or decreasing the heavy Higgs-boson masses, while still maintaining the possibility of sizable heavy Higgs-to-electroweakino decay rates, leads to scenarios where both search strategies are sensitive and complementary.\footnote{A detailed analysis and comparison of the LHC sensitivity to direct electroweakino production and to heavy Higgs-to-electroweakino decays within these (and other) scenarios is left for future work.}

In the light of the fact that the choice of the electroweakino-mass parameters of the \ltblchi scenario may already be in conflict with present \lhc{} data (if a dedicated analyses was performed), we want to close this section with some remarks on alternative scenarios with slightly heavier electroweakinos. In Tabs.~\ref{tab:EWino_scenarios1} and~\ref{tab:EWino_scenarios2} we also include detailed information on two alternative parameter points, scenario~3 and scenario~4, in which $M_1$, $M_2$ and $\mu$ are increased from their \ltblchi scenario values by $100$\,GeV. Again, the scenarios are distinct in the choice of $\tb$ values, $2.5$ and $7.5$, respectively. With respect to the original \ltblchi parameter points, scenario 1 and scenario 2, all electroweakino masses are simply increased by around $100$\,GeV. However, interestingly, there are no significant changes in the electroweakino and heavy Higgs-boson decay spectra, and our above discussion of collider signatures still holds.
Also the effect of raising the electroweakino spectrum on the Higgs-boson production cross sections is below the quoted precision.
On the other hand, current constraints from direct electroweakino searches are safely avoided for this alternative choice of $M_1$, $M_2$ and $\mu$, as the production rates for direct neutralino and chargino production are significantly smaller than for the original parameter choice, see \tab{tab:EWino_scenarios1}.
Lastly, note that due to the increased chargino masses, the enhancement of the $h\to \gamma\gamma$ rate is less pronounced in this case, as shown in the last row of \tab{tab:EWino_scenarios2}.


\subsection{Comparison with the hMSSM approach}

\begin{figure}[t]
\begin{center}
\includegraphics[width=0.49\textwidth]{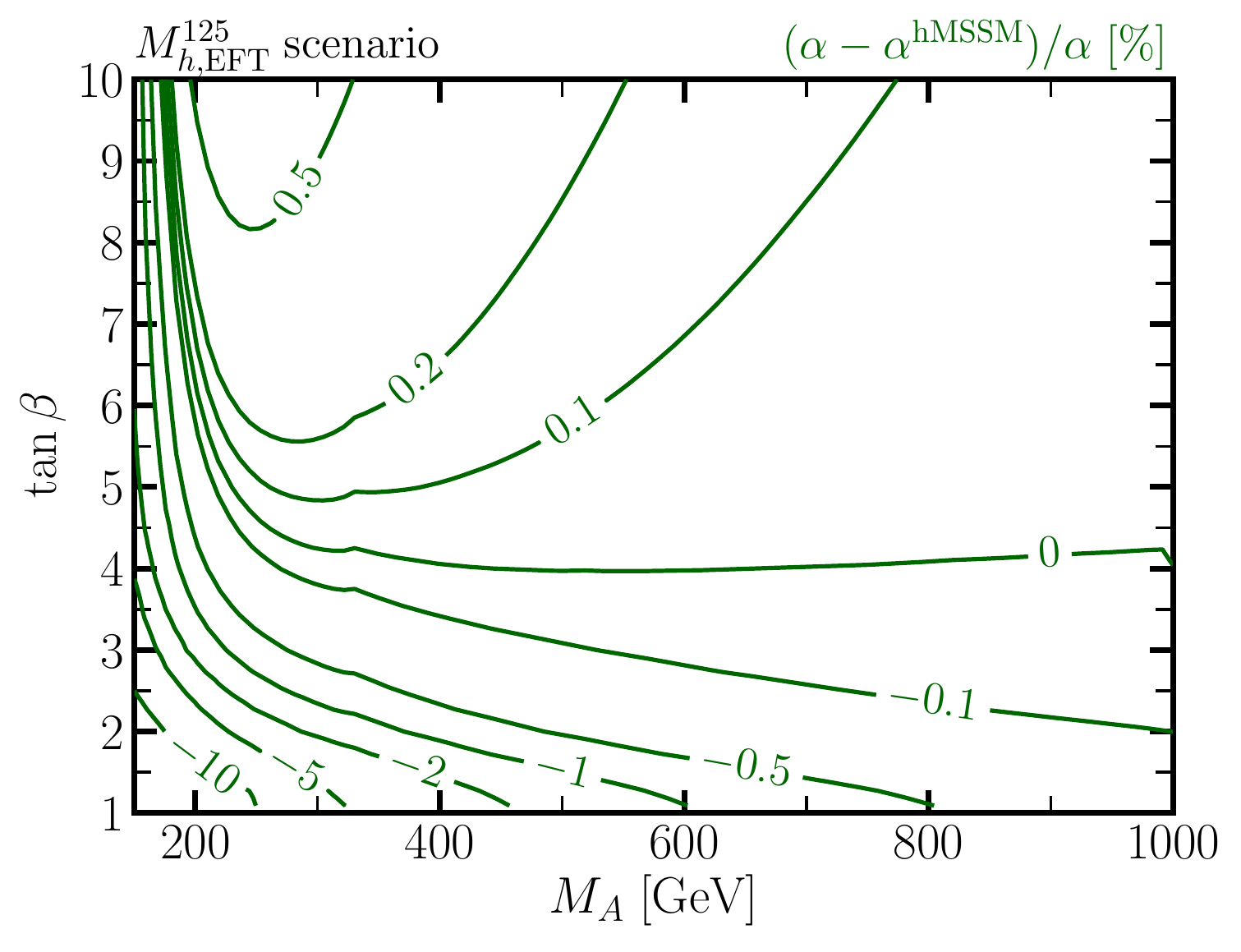}\hfill
\includegraphics[width=0.49\textwidth]{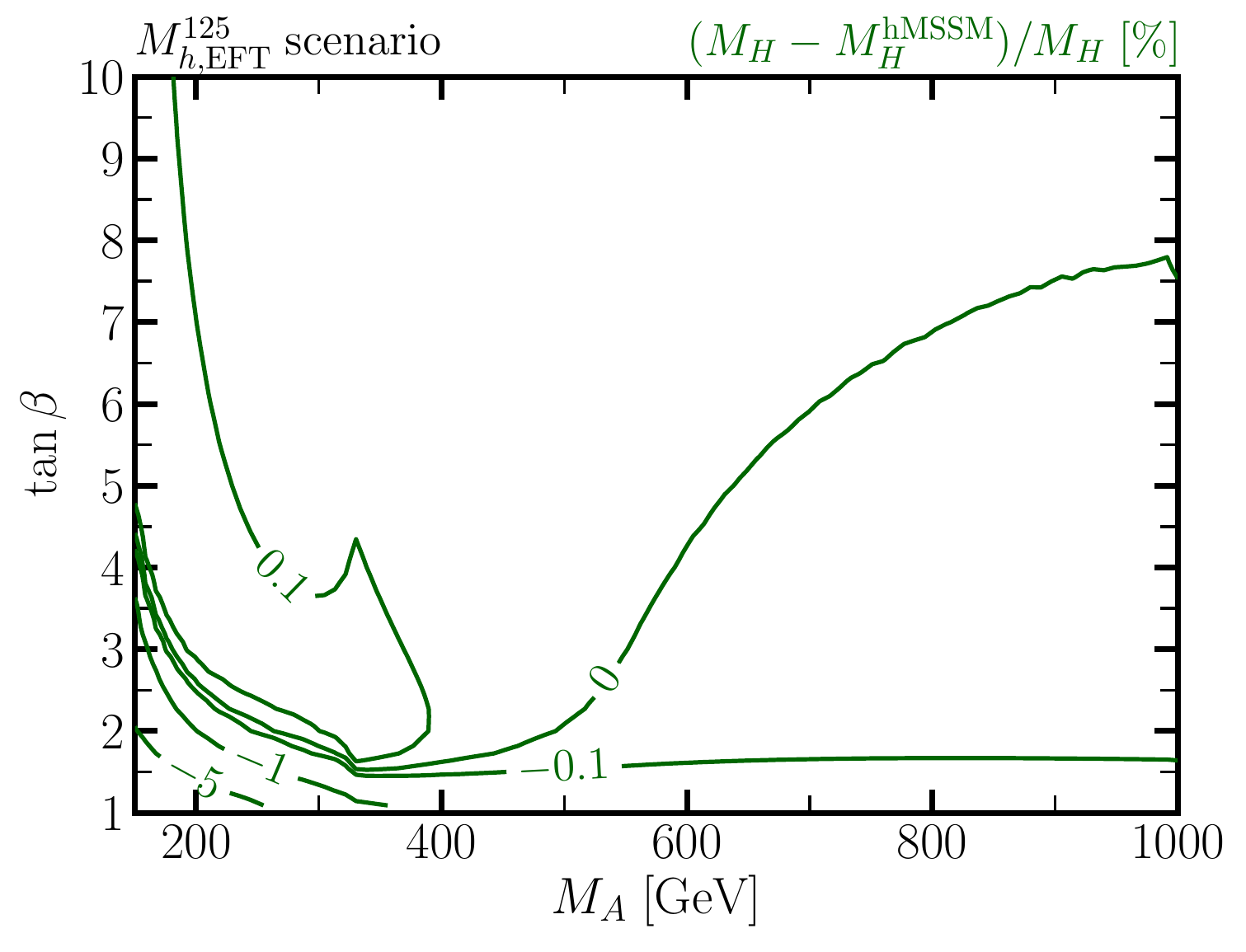}
\end{center}
\vspace*{-10mm}
\caption{\emph{Left}:
Relative difference in the prediction of the (effective) \cp{}-even Higgs-boson mixing angle $\alpha$ between the \ltbmhsc\ scenario
and the hMSSM approach in $\%$.
\emph{Right}: Relative difference in the prediction of the heavy \cp{}-even Higgs-boson mass $M_H$ between the \ltbmhsc\ scenario
and the hMSSM approach in $\%$.}
\label{fig:mh125-ltb_hMSSM}
\end{figure}

We already emphasized that the \ltbmhsc scenario is the perfect candidate to
assess the region of validity of the hMSSM approach~\cite{Djouadi:2013vqa,Djouadi:2013uqa,Djouadi:2015jea}, as it covers the same region in the ($M_A, \tan\beta$) parameter plane,
and for low values of $\mu/\msusy{}$ fulfills the hMSSM assumptions.
As a first step, we present a comparison of the predicted heavy \cp{}-even Higgs-boson mass, $M_H$, and the (effective) \cp{}-even Higgs-boson mixing angle, $\alpha$,
in \fig{fig:mh125-ltb_hMSSM}.\footnote{We leave a discussion of more elaborate quantities such as the Higgs-boson self-couplings and Higgs-to-Higgs decays to future work.}
 We only depict the range $M_A\in [150,1000]$\,GeV,
since at values of $M_A\lesssim 150$\,GeV the hMSSM approach is ill-defined --- the light-Higgs boson is not \sm{}-like ---
and for values of $M_A>1$\,TeV the observed differences are negligible.

According to \fig{fig:mh125-ltb_hMSSM} the hMSSM approach provides a very good approximation to both
the \cp{}-even Higgs-boson mixing angle and the heavy Higgs-boson mass at sufficiently large values of $M_A$ and $\tb$,
where differences are at the permille level.
On the other hand, at low values of $M_A\lesssim 600$\,GeV and $\tb\lesssim 4$, the discrepancies can reach a few percent
and at very low values, i.e.\ in the lower-left corner, even exceed $10$\%,
in particular for the (effective) \cp{}-even Higgs-boson mixing angle.
Such discrepancies are slightly larger than the differences observed between the
``low-$\tb$-high'' scenario and the hMSSM~\cite{Bagnaschi:2039911,Lee:2015uza}, where however lower values of
$\msusy{}$ were used in the low $M_A$ and low $\tb$ region.
In the experimentally allowed region with $M_A\gtrsim 700$\,GeV, however, the differences are well below the percent level
and thus the hMSSM approach provides a decent description of the \mssm{} Higgs-boson sector, at least within
its region of validity, i.e. for not too large values of $\tb$.

The hMSSM approach assumes that the ratio $\mu/\msusy{}$ is small. It therefore also makes sense to compare the hMSSM
approach against the \ltblchi scenario, which by definition
has an even smaller value of $\mu$ than the \ltbmhsc scenario. On the other hand, the \ltblchi scenario clearly violates one assumption of the hMSSM approach, namely
it comes with light electroweakinos, which alter Higgs-boson phenomenology substantially, as discussed in detail in the previous section.
Indeed, we find slightly larger differences in the prediction of the \cp{}-even Higgs-boson mixing angle and the heavy Higgs-boson mass
than those depicted in \fig{fig:mh125-ltb_hMSSM}, which are due to the Feynman-diagrammatic corrections induced by light electroweakinos on the
Higgs-boson self energies. However again the differences are well below the permille level in the experimentally still allowed region.


\section{Summary}
\label{sec:summary}
In this paper we have proposed two new benchmark scenarios for \mssm{} Higgs-boson searches at the \lhc{}, supplementing the scenarios suggested in \citere{Bahl:2018zmf}. In the scenarios proposed in \citere{Bahl:2018zmf} all \susy{} particles are below or close to the TeV scale. Consequently, the parameter region of $\tb\lesssim 8$ is incompatible with observations due to a too-low prediction of the \sm{}-like Higgs-boson mass. In this work we re-opened this parameter region by allowing for squark masses of up to $10^{16}$\,GeV, thus making it possible to reach a \sm{}-like Higgs-boson mass of $\sim 125$\,GeV even for low values of $\tb$ and $M_A$ (except for a region of very low values of $M_A< 200$\,GeV). The presented scenarios are designed to provide guidance for experimental efforts to probe the low $\tb$ region of the \mssm{} Higgs sector and also to motivate new \lhc{} searches for additional heavy Higgs bosons.

Our first scenario, the ``\ltbmhsc'' benchmark scenario, can be considered as extension of the \mhsc scenario~\cite{Bahl:2018zmf} to low $\tb$ values. In this scenario all supersymmetric particles have masses around or above the TeV scale. Consequently, the phenomenology resembles the one of a type-II \thdm{} with the Higgs-boson couplings constrained to be as in the \mssm{}. The strongest constraint in this scenario originates from the signal-strength measurements of the \sm{}-like Higgs boson, excluding the region of $M_A\lesssim 650$\,GeV. Since the \ltbmhsc scenario fulfills the assumptions of the hMSSM approach, it is a candidate for more detailed comparisons between a complete \mssm{} scenario and the hMSSM approach.
We presented a comparison for the predictions of the \cp{}-even Higgs-boson mixing angle and the heavy Higgs-boson mass and in the experimentally allowed region find discrepancies only at the permille level.
On the other hand, in particular the Higgs-boson self-couplings and Higgs-to-Higgs decays need further investigations. Since they are however hardly of relevance for our work, except from small corners of the parameter space that are already ruled out by Higgs-boson signal-strength measurements, we leave them to future work.

In our second scenario, the ``\ltblchi'' benchmark scenario, neutralinos and charginos are chosen to be light. This scenario represents an extension of the \lchi scenario~\cite{Bahl:2018zmf} to low $\tb$ values. The effect of low-mass charginos enhances the decay rate of the \sm{}-like Higgs boson into photons, in particular in compressed scenarios with large gaugino-Higgsino mixing. Future precision measurements of this rate will therefore indirectly probe a significant part of the parameter space with light electroweakinos. We furthermore studied in detail the possible decays of the heavy Higgs bosons into electroweakinos. While the presence of these decay modes weakens the sensitivity of \lhc{} searches for heavy Higgs bosons decaying into \sm{} particles, they also provide an interesting and promising new avenue for new physics searches. In fact, a signal in these channels would simultaneously reveal the presence of \bsm{} Higgs bosons \emph{and} supersymmetric particles. In particular, in scenarios with a compressed electroweakino mass spectrum (as chosen here), these signatures often feature multiple $W$- and/or $Z$-bosons, giving rise to multi-lepton final states, plus missing transverse energy. Moreover, as these electroweakinos originate from the decay of a heavy resonance, they can have sizable initial momentum, leading to better prospects for the reconstruction of leptons, as opposed to direct electroweakino production. \lhc{} searches for heavy Higgs bosons decaying to electroweakinos are therefore highly complementary to existing searches for direct electroweakino production. One of the main purposes of the \ltblchi benchmark scenario (as well as the \lchi scenario~\cite{Bahl:2018zmf}) is to motivate and initiate the design of dedicated searches for heavy Higgs-to-electroweakino decay signatures.

\section*{Acknowledgments}
We thank Emanuele Bagnaschi, Sven Heinemeyer, Gabriel Lee, Pietro Slavich, Alexander Voigt, Georg Weiglein and Carlos Wagner for helpful discussions.
This work was initiated in the context of the activities of the \lhc{} Higgs Cross Section Working Group (\lhchxswg).

{\footnotesize
\bibliographystyle{utphys}
\bibliography{MSSM_lowTB_benchmarks}
}

\end{document}